\documentstyle[12pt,epsfig,here,rotate,osamat]{retema}
\renewcommand{\baselinestretch}{1.5}
\renewcommand{\arraystretch}{1.5}
\begin{document}
\newcommand{\beq}[1]{\begin{equation}\label{#1}}
\newcommand{\eeq}{\end{equation}}
\newcommand{\bsq}[1]{\begin{subequations}\label{#1}}
\newcommand{\esq}{\end{subequations}}
\newcommand{\bfn}[1]{{\bf#1}}
\newcommand{\eqn}[1]{eqn. (I.#1)}
\newcommand{\eqns}[2]{eqns. (I.#1) and (I.#2)}
\newcommand{\gl}[1]{eqn. (\ref{#1})}
\newcommand{\gls}[2]{eqns. (\ref{#1}) and (\ref{#2})}
\newcommand{\fur}{\qquad\mbox{for }\,}
\newcommand{\taud}{$\tau_{\scriptscriptstyle D}$\ }
\newcommand{\mtaud}{\tau_{\scriptscriptstyle D} }
\newcommand{\taur}{$\tau^{\scriptscriptstyle R}$\ }
\newcommand{\mtaur}{\tau^{\scriptscriptstyle R} }
\newcommand{\taue}{$\tau^\varepsilon$\ }
\newcommand{\mtaue}{\tau^\varepsilon }
\title{Polymer--Mode--Coupling Theory of 
Finite--Size--Fluctuation Effects in Entangled
Solutions, Melts and Gels.  II. Comparison with Experiment}
\author{Matthias Fuchs$^+$
and Kenneth S. Schweizer\\
Departments of Materials Science and Engineering,\\  Chemistry,
and Materials Research Laboratory,\\
University of Illinois, 
1304 West Green Street, Urbana, Illinois 61801}

\date{May 1997}
\maketitle
\begin{abstract}
The predictions of the polymer mode coupling theory for  the finite size
corrections to the transport coefficients of entangled polymeric systems
are tested in comparisons with various
experimental data.  It is found that quantitative descriptions of the
viscosities, $\eta$, dielectric relaxation time, \taue, and diffusion coefficients, $D$, 
of polymer melts can be achieved with two microscopic structural
fit parameters whose values
are in the range expected from independent theoretical or experimental
information. An explanation for the (apparent) power law behaviors of $\eta$, \taue,
and  $D$ in (chemically distinct)
melts for intermediate molecular weights  as arising from finite size
corrections, mainly the self--consistent
constraint release mechanism, is given. 
The variation of tracer dielectric relaxation times from Rouse to
reptation--like behavior upon changes of the matrix molecular weight is
analyzed.
 Self and tracer
diffusion constants of entangled polymer solutions can be explained by the
theory as well, if one further parameter of the theory is adjusted.
The anomalous scaling of the tracer diffusion coefficients in semidilute and
concentrated polystyrene solutions, 
$D\sim N^{-2.5}$,
is predicted to arise due to the
spatial correlations of the entanglement constraints, termed ``constraint
porosity''.  Extensions of the theory to polymer tracer diffusion through 
polyvinylmethylether and polyacrylamide gels
provide an explanation of
 the observation of anomalously high molecular weight  scaling exponents
in a range where the size of the tracer, $R_g$, already considerably exceeds the
gel pore size, $\xi_g$.
\end{abstract}

\rightline{Macromolecules, accepted (May 1997).}

\noindent$^+$ Current address: Physik-Department, Technische Universit\"at
M\"unchen, D-85747 Garching, Germany

\newpage

\section{Introduction}

The microscopic polymer mode coupling (PMC) theory connects the
dynamics of entangled polymeric systems to the underlying equilibrium liquid
structure. This description therefore naturally includes 
corrections arising from the  finite molecular weights 
of tracer or matrix polymers. In the preceding paper \cite{pap1},
 referred to as paper I,  the PMC predictions with finite size corrections
have been worked out
for the transport properties of entangled solutions and melts, and for
polymer tracer motion in gels. In the present paper these
results are compared with data from many experiments. As the PMC approach
has been detailed in the preceding paper \cite{pap1}, 
it shall be summarized only shortly
 in this section and then compared to alternative approaches. The resulting
 formulae of paper I 
will be referenced by the equation numbers of  that paper with a 
prefix I. 

The PMC approach of paper I treats the Rouse--entangled crossover problem in a
simple interpolative manner.
The PMC contributions capturing
the entanglement corrections are combined with the Rouse, unentangled
quantities, see \eqns{52}{68} for $D$, \eqns{55}{67} for $\eta$,
and \eqns{58}{70} for \taue. Appreciable deviations from the reptation--like
asymptotes for high molecular weights are therefore not explained by
corrections of the Rouse dynamics to the asymptotic, entangled motion.
 In stark contrast to the contour length
fluctuation model of Doi \cite{doicont1,doicont2},  the repton model of
Rubinstein \cite{rubinstein},
and  the Rouse--fluctuating--chain--end calculations of O'Connors and Ball
\cite {oconnor}, the 
discrete nature of the tracer chain or its chain ends are not the cause of 
the deviations from the asymptotic molecular weight scalings. 

Finite size corrections to the transport coefficients within PMC theory result from
the  spatial and temporal correlations of the matrix constraints, called
entanglements. These constraints are modeled with a priori reasonable
assumptions. First, the local density variations enter through a
non--homogeneous compressibility, \eqn{45}.
Second, only a finite amplitude of the density fluctuations, or of the shear
stress amplitudes ( see section 3.B of paper I)
contributes to the elastic mesh formed by the entanglements. This amplitude
is modeled in analogy to  the entanglement plateau observed by neutron
scattering from single polymers \cite{richter1,richter2,richter3}.
 Except for an overall (small) amplitude, it is
characterized by the entanglement length, $b$, \eqn{46}.
These two effects are summarily termed ``constraint porosity''. Further, 
the temporal decay of the matrix constraints during the disentanglement step of
the matrix, see \eqns{47}{48}, is considered self--consistently; this 
process is termed
``constraint   release mechanism'' and provides another parallel channel for
the relaxation of the constraining excluded volume forces. 

Neglecting the constraint release and constraint porosity effects, PMC theory predicts
smooth crossovers from Rouse to entangled dynamics, as can be seen from eqns.
(I.52), (I.55), and (I.58). Reptation--like molecular weight scaling exponents
are observed. Still, it is important to note that these results
qualitatively differ from the corresponding reptation/ tube results
in the prediction of non--universal asymptotic prefactors. 
The effects on Markovian transport coefficients
of the internal mode  Rouse dynamics for finite tracer
molecular weights, which are the topic of the contour fluctuation, repton and 
chain--end fluctuation models \cite{doicont1,doicont2,rubinstein,oconnor}
for finite $N$ corrections to the terminal relaxation time, 
are neglected within the PMC approach of paper I. Recent dielectric
measurements of Adachi and coworkers \cite{adachi,adachi96} convincingly 
support the latter approach. A change in the molecular weight dependence of the
end--to--end--vector relaxation time of a tracer polymer is found upon  changes
of the molecular weight of the matrix polymers \cite{adachi,adachi96};
reptation--like scaling, $\mtaue\sim M^3$, is observed in immobile matrices of
high molecular weight polymers.
 This effect clearly cannot be explained by single polymer approaches
but must result from considerations of the dynamics of the matrix constraints.

The importance of the constraint release contributions in the viscosity  has
also been stressed by a recent simulation which focuses  on 
the configurational rearrangements of the entanglement points \cite{termonia}.
The consideration of the
dynamics of the surrounding matrix is an aspect of PMC theory it shares with the
constraint release models of Grassley \cite{grassley} and Klein
\cite{klein1,klein2}. 
Whereas in these  two approaches, the diffusion
coefficients of the reptative and constraint release dynamics are
phenomenologically added, in the  PMC approach the effects of both decay channels 
on the entanglement friction functions  depend on length scale and 
are derived from one set of
approximations for the microscopic intermolecular forces, as shown in
eqns. (I.11) to (I.27) of paper I. Moreover, the argument of self--consistently
determining the matrix disentanglement time from the dynamics of a single
matrix polymer is not included in the former phenomenological 
constraint release formulations,
whereas in PMC theory it is captured via  eqns. (I.64) to (I.66).
It appears that the plasticization model \cite{oconnor}, or the double
reptation formalism of Des Cloizeaux \cite{dobrep1,dobrep2,dobrep3,dobrep4},
can be viewed as a phenomenological attempt to incorporate these 
effects. In both models, the decay of the conformational dynamics or stress is
accelerated by taking a power of a single polymer correlation function.
From the perspective of
mode coupling approximations, non--integer powers as  advocated in
Ref. 5
seem difficult to motivate microscopically. 

In des Cloizeaux's work, additional phenomenological concepts such as ``time--dependent
reptation'' or ``a displacement dependent diffusion constant''
lead to the appearance of a new intermediate time scale of rather unclear
physical origin. This elaboration of the tube model
results in a 
 stronger--than--reptation scaling of the internal
relaxation time which
is tightly connected to the anomalous frequency dependence of
the high frequency wing of the disentanglement process in the shear modulus
\cite{dobrep2,dobrep3}. In experimental studies of shear moduli
\cite{baumg,jackwint,kannaan} and
dielectric spectra \cite{adachi1,adachi2,adachi3},
the high frequency wing of the disentanglement process consistently 
exhibits a power law
frequency dependence, $G''(\omega\mtaud\gg1)\sim \omega^{-x}$ and
$\varepsilon''(\omega\mtaud\gg1)\sim \omega^{-x'}$, with exponents clearly
much smaller than the reptation prediction, $x, x' = 0.2$ --- $0.3< 1/2$.  
In this respect it is an important finding of the
dielectric measurements  that the initial
decay of the disentanglement process is still more rapid, and consequently its
high frequency  wing still possesses a more shallow slope than the reptation
prediction, even in the limit where a reptation--like mass scaling of the
internal 
relaxation time is observed \cite{adachi,adachi96}.
Therefore,  we believe a consistent and convincing
physical  mechanism for the anomalous frequency
dependence of the initial decay of the disentanglement process remains
 to be
identified theoretically, and it  must be different from the mechanism leading to
non--reptation--like scaling of the longest relaxation time or viscosity.
 In PMC theory, this mechanism is the tracer shape fluctuations, as has been
shown in section 3.A of paper I, and will
further be discussed in section 3.E where the dielectric
measurements of Adachi and coworkers \cite{adachi} are analyzed; see also
Refs. 25---27.


Whereas various different non--selfconsistent phenomenological formulations of
 the constraint release mechanism \cite{grassley,klein1,klein2,watanabe}
have been considered as extensions of
the reptation/ tube approach, the constraint porosity effects of the PMC
description,  captured
in e.g. the \eqns{68}{69} 
have no analog in the phenomenological models for polymer melts and
solutions. This follows since these models  do not relate the entanglement
constraints to the static structure of the polymeric liquid. The 
measurements of tracer diffusion constants in strongly entangled
polymer solutions  \cite{nemoto1,nemoto2,nemoto3,nemoto}
find anomalously strong molecular weight dependences, e.g.
$D\propto N^{-2.5}$.
This  rules out the constraint
release mechanism as the only source of deviations from the reptation--like
scaling. Within the reptation/ tube approach only an extension of the repton
model, i.e. the numerical cage model of Deutsch and Madden
\cite{deutschmadden},  
claims to predict  molecular weight scaling of the diffusion coefficients and
viscosities  in agreement with the measurements in solution \cite{nemoto}.
However, from this phenomenological  model it is not apparent why such a
$M$--scaling is not observed in melts. 
 Moreover, this model would again fail in
explaining the reptation--like scaling of the internal relaxation time in the
limit of immobile matrices, as observed by dielectric spectroscopy
\cite{adachi,adachi96}.  
A similar concern applies to the ideas
 of Rubinstein and  Obukhov  \cite{obukov} where a close connection of the apparent
power laws of the diffusion coefficient and viscosity,
$D\eta\propto R_g^2$, is postulated for
solutions but not for melts.
Further non--reptation--like theoretical approaches
exist \cite{fixmann,douglas,herman1,herman2,loring1,loring2,avik},
  but generally 
have not been worked out in such detail
to allow comparison with the mentioned experimental observations and the issues
of finite $N$ corrections and apparent crossover scaling laws.

The static correlations of the matrix constraints as described by PMC theory naturally
generalize to amorphous media, especially gels, which are fractal on
intermediate length scales. In the context of polymer motion through fractal
media  it has been recognized by Muthukumar and
Baumg\"artner that strong deviations from the reptation--like behavior result
when the medium is characterized by a length scale or mesh 
of the order of the size of
the diffusing polymer \cite{baummu1,baummu2,baummu3,muthubaum}.
The entropic barrier model thus agrees with PMC theory
 in predicting finite size
corrections if the spatial length scale of the surrounding medium is not much
smaller than the radius of gyration of the tracer. 
However, in PMC theory the physically reasonable but difficult to quantify
 concepts, such as  ``entropic traps'' and 
``lack of topological correlations'' \cite{muthubaum} can be avoided, as the
tracer dynamics is directly  connected to the static structure
of the gel and to the tracer--gel interactions.
 Also, the limits of small and infinitely large tracer sizes and smooth
 crossovers between regimes are naturally
included in the PMC description but not in the entropic barrier model.
In the PMC approach,
the gel is assumed to exhibit pores of any (mesoscopic) size
smaller than the gel length scale, $\xi_g$, which therefore determines the
range of the spatial structural correlations. If flexible or non--rigid gels
structures are considered, then another length characterizing the elastic
stresses enters the PMC description, i.e. the \eqns{79}{80}.
In contrast with the entropic barrier \cite{muthubaum} considerations and its
enthalpic trap generalizations \cite{lumpkin} the PMC description of polymer transport 
through gels predicts strong finite size corrections to tracer diffusion even
if $R_g \gg \xi_g$. 

In the present paper the PMC results of paper I shall be compared to
experimental data in order to test the theoretical predictions.   A priori
estimates of the two non--universal parameters determining the finite size
corrections to the PMC dynamics are
presented in section 2.  Section 3 introduces the experimental systems and
studies examples for melt viscosities (sections 3.B and 3.C), melt self
diffusion coefficients (section 3.C and 3.D), melt tracer diffusion constants 
(section 3.D),  dielectric relaxation times in melts
(section 3.E), transport coefficients in solutions (section 3.F), and tracer
mobilities in gels (section 4). The paper concludes in section 5 with
 a summary of the results of this and the
preceding paper I, and suggestions for future experiments to further test our
predictions. 

\section{Estimates of Parameters}

The PMC theory of paper I predicts that the nonasymptotic corrections to the
dynamics of entangled polymers depend on two material and
thermodynamic--state--dependent dimensionless parameters of clear microscopic
meaning: $\delta=\xi_\rho/b$ and $\alpha$. The first parameter arises from
porosity or correlation effects which weaken the entanglement constraints. 
It involves two measurable quantities: the polymer density--density screening
length and the entanglement mesh length scale (``tube diameter'' in reptation
theory). Although such length scales can be estimated based on liquid state
integral equation theory or via computer simulations, the best source of
information is experiment. The second parameter enters the prefactor of the
asymptotic scaling of the diffusion constant with $N$, i.e. $D=D^R (N_e/N)
\lambda_D^{-1}$, where $D^R$ is the bare Rouse value and
$\lambda_D=32/3\alpha$. Thus, $\alpha$ quantifies an effective crossover
degree of polymerization for the  asymptotic behavior of $\beta D\zeta_0 \to
N_D/ N^2$ where $N_D=(3\alpha/32) N_e$. A microscopic connection with the mean
square force exerted on the probe chain center--of--mass by all the surrounding
polymers has been derived: $\lambda_D=32/3\alpha \propto \langle |F|^2 \rangle
/\varrho_m$, where $\varrho_m$ is the segmental number density. Hence, $\alpha$
is inversely proportional to the strength of the constraining forces
(entanglements) per unit matrix number density. Based on liquid state theory,
and the Kirkwood superposition approximation, one can derive \cite{ks3,ks4}:
$\langle |F|^2 \rangle /\varrho_m\propto (g_d \xi_\rho)^2$, where $g_d$ is the
contact value of the interchain segment--segment radial distribution function
which quantifies the local hard core excluded volume forces between chains. 
Although experimentally accessible in principle (and definitely via
simulations), $\langle |F|^2\rangle$ and $g_d$ do not appear measurable in
practice. Thus, we are forced to make a priori estimates of $\alpha$ based on
theoretical input. 

In this section, we discuss the available experimental information, and
theoretical estimates, of the required microscopic parameters. In section 3 we
treat them as  adjustable parameters determined by fits to  viscosity and/ or
diffusion constant data of specific polymer solutions and melts. The
independent, a priori estimates of this section provide strong constraints on
the physically acceptable values these parameters can assume. 

We also note that, in principle, the entanglement degree of polymerization,
$N_e$, and the polymer material parameters determining its bare (Rouse)
dynamics, are also required as input to the theory. However, we always consider
ratios of transport coefficients relative to the unentangled Rouse values,
$\eta/\eta^R$ and $D/D^R(N_e)$. Thus, we employ the PMC prediction \cite{pap1}
that these ratios depend on degree of polymerization only via $N/N_e$, where
$N_e$ is taken from experiment and data is plotted in the standard reduced
variable format. Thus, comparison to experiment requires only specifying the
two nonuniversal parameters $\delta$ and $\alpha$. 

\subsection{Experimental Estimates}

Consider first the ratio $\delta = \xi_\rho/b$. In the Table we list
rheologically--extracted (via the relation $G_N=\varrho_m k_B T(\sigma/b)^2$) 
values for the entanglement mesh, $b$, of several systems in the melt
\cite{fetters}. Values appropriate for solutions containing a fraction $\Phi$ of
polymer follow from the known scaling relations \cite{brownsc,de}:
$b\propto \Phi^{-\nu}$, where $\nu\approx$0.75 (good solvents where
$\xi_\rho\propto b$ and $\sigma\propto \Phi^{-1/8}$) or $\nu\approx2/3$ (theta
solvents \cite{colrub}). Remarkably, these scaling relations appear to apply
over a wide polymer concentration regime from semidilute up to the melt
\cite{pearson87}. The density screening length has been measured for only a few
polymers, and up to $\Phi\approx$ 0.3---0.4 at best. For polystyrene, the
average of many experiments yields \cite{brownsc}: $\xi_\rho\approx 2.5 \AA
\Phi^{-0.72}$ (good solvents) and $\xi_\rho\approx 6. \AA
\Phi^{-1.0}$ (theta solvents). Of course, as the melt state is approached,
theta and good solvents become equivalent, and hence these laws must merge into
a single concentrated solution behavior. Elementary dense liquid considerations
require that such simple scaling laws for the screening length cannot hold up
to the melt (where $\xi_\rho < \sigma$). The screening length (and hence the
osmotic pressure) decreases (increases) much more rapidly than in semidilute
solution and in a non--power law fashion \cite{kcur2,kcur3,thread,hansen}. Some
nonuniversality is expected in the concentrated and melt regime. 

Extrapolation of polystyrene semidilute good solvent scaling laws up to the
melt  suggests a screening length of roughly 3--5 $\AA$, which is
expected to be an upper bound. Interestingly, as is seen from the Table, this
extrapolation is in good accord with a value calculated based on the invariant 
packing length \cite{fetters}, $p=1/\varrho_m\sigma^2$. The packing length
rigorously emerges as essentially equal to $\xi_\rho=(\pi/3) p$ from the
simplest analytic version of PRISM theory based on the idealized Gaussian
thread or string models \cite{ks3,ks4,kcur2,kcur3,thread,kcur1}.
In any case, for polystyrene an upper bound for the parameter
$\delta=\xi_\rho/b \approx$ 0.05 in the melt can be estimated from the
experimental data. It is probably smaller, perhaps by a factor of 2 or 3. This
latter estimate follows from the thread PRISM based connection between the
density screening length and the dimensionless measure of long wavelength
density fluctuations \cite{ks3,ks4,kcur2,kcur3,thread,kcur1}:
$S_0 = \varrho_m k_BT \kappa = 12 (\xi_\rho/\sigma)^2$, where $\kappa$ is the
isothermal compressibility and $\sigma$ the statistical segment length. From
scattering experiments or thermodynamic measurements on dense polymer liquids,
one finds that $S_0$ is typically in the range of 0.15---0.3. Thus, for
polystyrene with a statistical segment length of 7 $\AA$\ one obtains a
$\xi_\rho\approx 1 \AA$. On the other hand, perhaps the dynamically--relevant
mesh size for entanglements involves only the backbone mass, which would again
result in an effective mesh size of 2---3 $\AA$ for polystyrene. 

In solution one expects different behaviors depending on solvent quality and
also chemical structure. In good solvents, where the screening length and
entanglement length are predicted (by scaling \cite{de} and PMC \cite{ks3}
theories) and measured to be proportional \cite{de,raspaud}, one expects a
polymer  concentration--{\it independent} value of $\delta$ but nonuniversality
with respect to the chemical structure dependence. The latter failure of naive
scaling has been emphasized by Adam and coworkers \cite{raspaud} who find a
correlation of this ratio with the melt $N_e$. For semidilute good polystyrene
solutions, from the arguments given above and known data we expect a larger
value of $\delta$ than in the melt, perhaps in the range of 0.2---0.3. Recent
Monte Carlo simulations \cite{paulbinder,binder} of semidilute good polymer
solutions of weakly entangled chains find a value of $\delta\approx$ 0.3. Even
larger values are expected in theta solvents since the screening length
increases faster than the entanglement length as polymer concentration is
decreased \cite{brownsc,colrub,pearson87}. Moreover, $\delta$ is expected to be
polymer concentration--{\it dependent} in theta solvents ($\delta \propto
\Phi^{-1/3}$ roughly) with increasing values for less concentrated polymer
solutions. 

Based on the above considerations, we have a rather good idea of the magnitude
of the key parameter $\delta$ in the melt, and in good and theta solutions, at
least for polystyrene. Although, we expect the concentration dependence of
$\delta$ is relatively universal, the magnitude may depend significantly on
chemical structure. Direct measurements of the density screening length in
other materials is required to determine just how large a variation is
possible. 

Estimation of the parameter $\alpha$, or equivalently $\lambda_D=32/3\alpha$,
from experimental data is more difficult. If the asymptotic scaling regime 
($D\propto N^{-2}$) is achieved, then $\alpha$ is immediately
determined. However, this appears possible only in melts, and not in
solution. The existing experimental data has clearly established that
$\lambda_D$ is significantly larger in semidilute and concentrated solutions
of polystyrene compared with its melt value
\cite{nemoto1,nemoto2,nemoto3,nemoto}. 

\subsection{Theoretical Estimates}

As discussed elsewhere \cite{ks3,ks4}, theoretical estimates of the density
screening length, entanglement mesh or $N_e$, $\langle |F|^2 \rangle
/\varrho_m$, and $g_d$ have been made based on a very simple chain model
(Gaussian thread or string), analytic PRISM theory, and the RR and PMC
dynamical theories. This analysis suggests $\langle |F|^2 \rangle
/\varrho_m$, or $\alpha$, is nearly polymer concentration and chemical
structure independent due to a nearly perfect inverse relationship between the
contact value $g_d$ and the density screening length. This work provides some
first principles justification for polymer density scalings observed experimentally,
but quantitatively accurate calculations are difficult. Estimates based on
numerical PRISM computations for solutions and melts which employ more
chemically realistic chain models (semiflexible Koyama or RIS \cite{ted}) 
produce results in surprisingly good semi--quantitative agreement with the
prior analytic work, even up to melt--like densities where $g_d^2S_0\approx$
0.1---0.2. Based on \eqn{54}, we obtain the crude ab initio estimate of
$\alpha\approx 2/g_d^2S_0\approx$ 10---20.

At a more detailed level, it is found that $\langle |F|^2 \rangle
/\varrho_m$ (proportional to $\alpha^{-1}$) decreases weakly with polymer
concentration, and is also (modestly) quantitatively sensitive to local
chemical features such as the effective chain aspect ratio \cite{ks3,ted}.
Since  $\langle |F|^2 \rangle /\varrho_m \propto (g_d\xi_\rho)^2$, this
behavior reflects the fact that the density and chemical structure dependences
of the interchain segment--segment contact value and density screening length
are not precisely inversely proportional to each other with a  universal
prefactor. This conclusion seems unavoidable to some degree in any
concentration regime where the physical requirement for scaling ideas to apply,
$\xi_\rho\gg \sigma$ or $d$, does not hold. Thus, based on these theoretical
studies, we expect some (modest) chemical structure variability of $\alpha$ (or
$\lambda_D$) and a decrease (increase) in magnitude as the melt is diluted with
solvent. 

As discussed previously \cite{pap1,fractal}, by comparing PMC theory with the
reptation/ tube theory in the asymptotic limit the parameter $\alpha$ is easily
shown to be approximately 3. Of course, this is only obviously relevant in
polymer melts since it is the only case where the asymptotic $N$--scaling of
the diffusion coefficient is observed. In the reptation/ tube picture, $\alpha$
is predicted to be universal, independent of chemical structure, polymer
concentration, and solvent quality. Such universality of $\alpha$ results in a
predicted concentration, solvent quality, and chemical structure {\it
independence} of the constraint release corrections for viscosity and terminal
relaxation time. Although not literally true for PMC/ PRISM theory, these
reptation/ tube predictions of a relative insensitivity of $\alpha$ seem in
rough accord with our microscopic approach. We note that computer simulations
have never attained the asymptotic regime, so they provide no knowledge of the
parameter $\alpha$. 

We mentioned some potential limitations of the equilibrium structural model
employed in paper I. The simple  Ornstein--Zernicke form adopted for $S_k$
(see \eqn{45}) is adequate for semidilute and moderately concentrated
solutions. However, as the melt state is approached the density screening
length becomes so short that it is not obvious if such a form
applies. However, the simple equilibrium model still seems justified since
experimentally the inequalities $R_g, b\gg \xi_\rho$ apply to entangled systems
\cite{colrub,raspaud,binder}, and hence we only need to know $S_k$ for
$k\xi_\rho\ll1 $. Another caveat is that we are treating a two component 
polymer--solvent mixture as an effective one--component system. Although this
is standard practice in coarse--grained polymer theories, it may be inadequate
for accurately computing the chemical structure, temperature, and
concentration dependence of the structural properties required by our
microscopic dynamical theory. This fact, along with our neglect of hydrodynamic
interactions, makes quantitatively--reliable a priori theoretical estimates of
the $\alpha$ parameter especially difficult in solution.

Finally, we point out that both $\alpha$ and $\delta$ will in general depend
(presumably weakly) on temperature via the implicit dependences of the relevant
length scales, density, and equilibrium mean square forces. 
\section{Comparison with Melt and Solution Experiments}

\subsection{Experimental Parameters}

In order to test the description of finite size effects within the PMC
theory  a number
of experimental data sets are studied. The shear viscosities of  melts of
polybutadiene \cite{colby}  (PBD) and hydrogenated polybutadiene
\cite{pearson}  (PBDh), and the melt self diffusion coefficients of 
\cite{fleischer} PBD, PBDh \cite{pearson}, polyisoprene \cite{fleischer} (PI), 
polydimethylsiloxane (PDMS), polyethyleneoxide  \cite{appel} (PEO), and of
polystyrene \cite{anton}  (PS) are considered. 
In order to study the importance of the constraint release mechanism in melts,
measurements of PS tracers in PS matrices of different molecular weights
\cite{greenkramer}, and relaxation times from dielectric spectroscopy of PI
tracers in PBD melts \cite{adachi}, and the neat PI melt case, 
 are analyzed as well. Self diffusion, $N=P$,  and   
infinite--matrix--molecular--weight--tracer diffusion coefficients, $P\gg N$,
for three different concentrations of polystyrene dissolved in dibutyl
phthalate, a good solvent, are also included \cite{nemoto1,nemoto2,nemoto3}.
Here and throughout the paper, $P$ denotes the molecular weight or degree of
polymerization of the matrix polymers, and $M$ or $N$ the tracer values,
respectively. Measurements extending to high reduced degrees of polymerization,
$n=N/N_e$, are of special interest. The measurements of Nemoto and coworkers on
semidilute and concentrated 
solutions of polystyrene are also included. At present, no convincing
theoretical
understanding of these data exists  within the reptation/ tube approach
\cite{nemoto}. 

Material parameters relevant to  the entangled polymer
dynamics are listed in the Table. The 
values of the molecular weights of entanglement, $M_e$, are the ones
reported in the quoted experimental studies and used in our analysis. 
From the neutron scattering and rheological data survey of Ref. 46
values for the entanglement length, $b$, and for the so--called packing length
$p$ (and estimated  density screening length
$\xi_\rho$) are included. Within Gaussian thread PRISM theory, $\xi_\rho$
is rigorously identified with the
packing length, $p$. The data
sets were shifted vertically on logarithmic scale in order to determine the
prefactors $k_BT/\zeta_0N_e$ or $\eta^R$ which lead to overlap with the Rouse
result, $D\zeta_0N_e/k_BT \to 1/n$ and $\eta/\eta^R \to 1$, respectively, for
small molecular weights, $n=N/N_e=M/M_e\ll 1$. Note that in all cases data for
$n\le 1$ were available, and as little arbitrariness was introduced in the
shifts as possible. The PMC dynamical  parameters, $\alpha$ and $\delta$,
listed in the Table  follow from the fits discussed  in the following sections,
and carry significant uncertainties.

\subsection{Melt Viscosity}

A PMC analysis of the reduced viscosity, $\eta/\eta^R$, versus reduced
molecular weight, $n=N/N_e$, highlights the effects of the self--consistent
 constraint release
mechanism and involves one fit parameter only, the inverse strength parameter
$\alpha$. Comparing the asymptotic reptation and PMC predictions for
the viscosity at $n\to\infty$, the universal  value $\alpha\approx 3.2$ is
estimated. Figure \ref{fpbd} shows the viscosities of PBD melts for which the
highest experimental 
reduced molecular weights have been reported \cite{colby}. The
theoretical result, eqn. (I.74), has been used. The PMC result (at least)
 semi quantitatively accounts for the measured data for all
molecular weights. The asymptotic scaling, $\eta\sim N^3$ (not shown in figure
\ref{fpbd}), is not yet exhibited by the data nor by the PMC fit even though
values $n=10^4$ are achieved. The matrix constraints decay on the same time
scale which determines the viscosity, i.e  the disentanglement time, \taud, and
less friction therefore arises than in the asymptotic limit of frozen
entanglement constraints. The constraint porosity effects, i.e. the spatial
correlations of the entanglements characterized by nonzero values of
the density screening
length, $\xi_\rho$, and the entanglement length, $b$, have been neglected in
figure \ref{fpbd} as they do  not noticeably affect the shape of the
theoretical curve for expected values of the length scale ratio in the melt,
$\delta=\xi_\rho/b\approx 1/20$.  

A second fit of the PMC results to the data is shown in figure \ref{fpbd}. It
assumes, as has been pointed out by O'Connor and Ball \cite{oconnor}, 
that the monomeric friction coefficient, $\zeta_0$, found in the Rouse regime
in Ref. 61
 is somewhat too high. O'Connor and Ball argue by
comparing the friction coefficients from Ref. 61
 with an earlier
measurement by Roovers \cite{rooverpbd}, that the Rouse results of Ref. 61
are a factor 2.346 too high, $ \zeta_0^{\rm CFG} = 2.346 \zeta_0$. If this
correction of the monomeric friction coefficient is taken into account, by
fitting the expression $\eta/\eta^R=1+\frac{\lambda_\eta}{2.364} \beta n^2$ to
the data instead of \eqn{74}, then a quantitative fit of the PMC results to the
measured viscosities becomes possible in the full range of molecular weights
excluding the highest data point only. The value found for the strength
parameter $\alpha$, $\alpha\approx 7.7$ without, and $\alpha\approx 4.9$ with,
this correction, then lies closer to the reptation prediction of $\alpha\approx
3$ and also to
values  found in other polymer melts, e.g. PBDh (see below).
Such values of
$\alpha$ are also  expected from the
observation of the apparent power $\eta\sim N^{3.4}$, as demonstrated in the
model calculations shown in section 5.A of paper I, where a power $\eta\sim
N^{3.4}$ is observed for values of $\alpha$ below 5. 

After correction of the monomeric friction coefficient, all data except for the
highest molecular weight point (which is not reliable \cite{plazek}) 
fall close to the PMC fit. This is especially
apparent from the inset in figure \ref{fpbd} where the relative variation of
the viscosity compared to its asymptotic behavior, $\eta/M^3$, is shown. The
data scatter rather unsystematically around the fit with deviations well
compatible with the error bars reported in Ref. 61
 due to
uncertainties in the measurements of the molecular weight and the viscosity.  
The slow approach of the constraint release corrections to the asymptote,
\eqn{69}, explains why the asymptotic $\eta/M^3\to$ const. is not yet
observed. 

From figure \ref{fpbd}, and especially from the inset, it is suggested that
the apparent power law behavior of the viscosity in an intermediate molecular
weight range, $\eta\sim N^{3.4}$, is not the sign of a true power law scaling
behavior but arises from the competition of a finite size effect, namely the
self--consistent
constraint release mechanism, and the asymptotic, reptation--like scaling,
$\eta\sim M^3$. Clear observations of the true asymptotic power are not
expected  in the experimentally accessible molecular weight range, as the
constraint release correction factor, $\beta(n/\alpha^2)$, in \eqn{67}, 
shows an
extremely slow approach to unity, $1-\beta\to 4/3 (\alpha^2/n)^{1/4}$
\eqn{73}. Effectively, the slow variation  of $\beta$ around $n\approx 10^3$
may mimic the asymptote, \eqn{54}, which, however, correctly describes 
the data for much higher rescaled molecular weights $n$ only.

Unfortunately, no diffusion data exist for PBD melts  out to such high
molecular weights as used for the viscosity measurements. In the next section,
the combined comparison of the theory with diffusion and viscosity data
provides a further more stringent test of the physical picture of PMC theory.

\subsection{Combined Viscosity and Self Diffusion Melt Data}

Of special value to test the close connection of the center--of--mass and
conformational dynamics as it is predicted by  PMC theory are data on both dynamics
for identical polymer systems. An interesting example are the measurements in
PBDh by Pearson and coworkers which extend to rather high degrees of
polymerization and provide self diffusion constants and viscosities for the
identical samples \cite{pearson}. The viscosity analysis, using the PMC
result, eqns. (I.74), fixes the inverse strength parameter $\alpha$
and therefore the constraint release contribution to $\eta$ and $D_s$.
The value $\alpha=5.2$ gives the best least square fit to the viscosity data
and is rather close to the value fitted to the PBD viscosities.
Figure \ref{fpbdh} displays the data and PMC results. A universal
value of $\alpha$ is not expected by PMC theory; the magnitude agrees well with the
PRISM estimate and the universal value predicted by reptation theory. The
apparent power law behavior of the viscosity \cite{pearson},
$\eta\sim N^{3.4}$, is again explained as a finite size crossover effect due
to the  constraint release mechanism. 

A slightly smaller parameter, $\alpha=4.3$,
provides the combined best PMC fit to the viscosity and the self diffusion
coefficient as a function of reduced molecular weight. The lower part of figure
\ref{fpbdh} compares the two viscosity fits which exhibit similar fit quality.
 In
the upper part of figure \ref{fpbdh} the self diffusion coefficients are seen
to overshoot slightly compared to the reptation like asymptote, $D\to D^R/
\lambda_D n$, \eqn{52}. This small enhancement of the diffusion constant
again results from the constraint release mechanism as is shown by the PMC fit
using eqns. (I.64,I.68,I.69) and 
the same parameter, $\alpha=4.3$, as the viscosity fit. The small density
screening length, $\delta=\xi_\rho/b\approx0.03$, found in this analysis
 indicates
small constraint porosity contributions to $D$. 
Moreover, this value for $\delta$ is in good accord with the independently
measured  length scales listed in the Table.
An independent least square
fit to the diffusivities alone leads to a slightly smaller value of 
$\alpha=3.5$, but somewhat a  larger constraint porosity contribution,
$\delta=0.04$. The rather small variations of the parameters,
$\alpha=4.3\pm1$ and $\delta=0.03\pm0.01$ for PBDh, however, are well
compatible with both
the simplifications of the polymer matrix model, \eqns{45}{46},
used in the theoretical approach, and a priori estimates.

 It is the decay of the matrix constraints which is the
dominant finite size effect in  polymeric melts. Therefore, the larger finite
size corrections in the viscosity compared to the diffusion coefficients are
expected  in PMC theory.
 In paper I, it was discussed that the larger intramolecular
weighting of the matrix entanglement contributions for long wavelengths
leads to larger constraint release effects in
the conformational dynamics than in the center--of--mass motion. The
observation in polymer melts of the asymptotic scaling, $D\sim N^{-2}$, in the
self and tracer (for $P>N$) diffusion coefficients, but not in the shear
viscosities, is
 therefore naturally  explained by the different intramolecular correlations
captured in the $\Sigma$ and $M$ memory functions of PMC theory.
 See sections 4
and 5, and figures 4 to 8, in paper I for more discussions of this connection.

\subsection{Melt Tracer and Self Diffusion Constants}

The small corrections of the finite size effects in the melt diffusion data
have two origins in PMC theory.
 First, the constraint release mechanism is rather
ineffective for the center--of--mass motion because
of the local nature of the $\Sigma$ memory function.  Second, in melts the
spatial correlations of the matrix constraints are characterized by rather
small lengths scales, $\xi_\rho$ and $b$, and a length scale ratio
$\delta=\xi_\rho/b$,  much smaller than in the solution cases.

 Self diffusion data
for different melts do not fall on a single master curve, $D/D^R = f(N/N_e)$,
as predicted by reptation, because of the variations in the (inverse)
entanglement strength factor, $\alpha$. Independent fits to a number of polymer
melt self diffusion constants (PBD, PBDh, PDMS, PEO, PI, PS),
however, show that values of $\alpha$ fall in a rather tight range of 
 $\alpha=2.5$ --- 5. The
density screening lengths, $\xi_\rho$,   and length scale ratios,
$\delta=\xi_\rho/b$, found in these fits lie consistently below $\delta=0.06$
as expected for dense melts (see the Table). 
Figure \ref{fmeltb} displays a representative set of self diffusion data and
corresponding PMC fits.

A particularly  interesting set of self
diffusion data is included in figure \ref{fmeltb}. Diffusion constants
of PDMS \cite{appel} exhibit an apparent power law, $D\sim N^{-2.4}$, up to ca.
$log(N/N_e)=1.2$. Such high apparent power law exponents for rather small
degrees of polymerization are  well 
rationalized by the PMC finite size effects, mainly the constraint release
mechanism, as shown by the PMC fit in figure \ref{fmeltb}. During
this fit three parameters, $\alpha$, $\lambda_D$ and $\delta$, were varied
freely in eqns.    (I.64,I.68,I.69). Supporting the close connection of the
center--of--mass and the conformational motion predicted by PMC theory and leading to
\eqn{53} for $\lambda_D$, the two resulting fit parameters, $\alpha$ and
$\lambda_D$,  almost satisfy \eqn{53}, $32/3\alpha\approx3.8$ compared to
$\lambda_D=3.6$.  

A  common PMC fit to a number of self diffusion data sets using
eqns.  (I.64,I.68,I.69) is shown in figure
\ref{fmelt}. Rather small enhancements of the self--diffusion 
constants compared to the asymptotic value
 result from the small consequences of the constraint release
mechanism in the melt.
 The small value of the extracted  $\delta$, $\delta=0.05$, indicates that
constraint porosity can be neglected in this fit.

Without showing a corresponding figure, let us report that for tracer polymers
in a strongly entangled polymer melt, where the matrix molecular weight far
exceeds the tracer molecular weight, $P\gg N$, 
 almost no deviations from the asymptotic
prediction, $D=D^R/(1+\lambda_D n)$, \eqn{52}, are observed in the PMC results
in agreement with experimental measurements \cite{greenkramer}.
In figure 7 of paper I, model calculations for melt--like 
parameters, $\alpha=3$
and $\delta\le 0.05$, show a smooth crossover of this infinite matrix molecular
weight tracer diffusion constant from its Rouse, $D\sim N^{-1}$, to
asymptotic, $D\sim N^{-2}$, behavior with no overshooting or higher apparent
exponents in intermediate $N$ regimes. This behavior clearly contrasts with the
rather strong overshooting or rather high apparent exponent, $D\sim N^{-2.4}$,
caused by the constraint release mechanism in the PDMS self diffusion data of
figure \ref{fmeltb}. Note that for $n=N/N_e=4.4$, the PDMS self diffusion
constant lies a factor 2.1 above the asymptote obtained by neglecting finites
size effects. The individual  analysis of the PBD and PEO self diffusion
coefficients in figure \ref{fmeltb} further support 
these findings; small length scale ratios are observed and dominating
constraint release corrections to the asymptotic behavior.
 
The importance of the dynamics of the matrix constraints on the tracer
diffusion coefficients can most comprehensively be studied from a set of data
for different tracer, $M$,  and matrix polymer, $P$,  molecular
weights. Recalling that our 
approach applies to entangled polymer matrices only, due to the approximations
in the RR model, we study the data of Green and Kramer \cite{greenkramer}
for PS tracer in PS melts for $P>P_e$. As the authors did not publish tables of
their extended set of data we use the reported fitted theoretical curves in
order to compare with 
the PMC results. Note that Green and Kramer could fit their data equally well
with the models of Grassley \cite{grassley} and Klein \cite{klein2}. Both
models mainly differ in the steepness of the apparent 
$P$ dependence of the tracer
diffusion constants  in the crossover range, $P\le N$.
 We use the fitted model results of Grassley because we expect easier
agreement of  PMC theory with the model by Klein; in figures 8 and 9 of paper I,
$P$--dependences around Klein's prediction, $D\sim P^{-2.5}$, are seen in the
PMC results. 
Grassley's expression  exhibits a stronger $P$--dependence,
\beq{grass}
D(M,P) \propto \frac{1}{M^2} ( 1 + \alpha_{cr} M_e^2 M P^{-3} ) \; .
\eeq
Note that \gl{grass}
does not reproduce the Rouse model for small tracer molecular weights, $M<M_e$.
Therefore we normalize it to unity for $M=M_e$, i.e. study $D(M,P)/D(M_e,P)$,
and do not require the connection of $\alpha$ and $\lambda_D$,  \eqn{53},
i.e. the relative ratio of Rouse to asymptotic PMC result, in the PMC fit. 
The experimental parameters, $M_e$ as in the Table
 and $\alpha_{cr}=\frac{48}{25}
z (\frac{12}{\pi^2})^{z-1}$ with $z=3.5$, are taken from Ref. 66
and therefore mimic the experimental data for the  PS tracer in PS melt
systems. The 
values for the tracer and matrix polymer molecular weights used in figure
\ref{fgrass} span the experimentally studied range.  Even
though the PMC results in figure \ref{fgrass} do not reproduce the $D\sim
P^{-3}$ prediction of Grassley for the tracer diffusion constant in the the
constraint release dominated region, overall a reasonable fit to Grassley's  
model can be achieved as shown. The value $\alpha=3$ is chosen a priori
as appropriate
for polymer melts, and the parameters $\delta=0.07$ and $\lambda_D=1.58$ are
found from least square fits.  Comparing the shapes of the PMC results with the
actual diffusion data of Ref. 66
 qualitative agreement can be
deduced.  It appears that the PMC theory
adequately describes the constraint release decay of the
matrix entanglement constraints as observed in tracer diffusion measurements in
entangled polymeric melts.  

\subsection{Dielectric Relaxation Times} 

An especially powerful technique to study the conformational tracer dynamics,
especially the end--to--end vector fluctuations, is the dielectric spectroscopy
of type--A tracer polymers in type--B polymer melts
\cite{adachi,adachi96}. Type--A polymers are defined to possess monomeric
dipole moments parallel to the chain contour, whereas type--B polymers possess
perpendicular moments only \cite{stockmayer}. Adachi and coworkers have studied
such a system, PI tracer in PBD melts, for varying tracer and matrix molecular
weights \cite{adachi}. They report the molecular weight dependence of the
tracer dielectric relaxation time and the high frequency asymptote of the
disentanglement process in the dielectric spectrum. It is important to recall
from paper I that PMC theory identifies two different mechanisms, one which is
responsible for the initial decay of the end--to--end vector correlation
function, $\langle {\bf P}(t) \cdot {\bf P}(0) \rangle / \langle {\bf P}(0)
\cdot {\bf P}(0) \rangle -1 \sim (t/N^y\tau_0)^x$, and two other ones, which
affect the dielectric relaxation time. At first, let us discuss the latter,
namely the constraint release and constraint porosity effects, and compare them
with the experimental data for \taue, the dielectric relaxation time.

As shown in section 5.D of paper I, PMC theory predicts three different asymptotic
behaviors for the relaxation time of the end--to--end vector correlation
function of a tracer polymer. In unentangled matrices, $p=P/P_e<1$, the Rouse
behavior, $\mtaue\sim N^2$, should be observed. In strongly entangled matrices,
where the surrounding polymers are immobile relative to the tracer, the
dielectric follows the internal disentanglement time, $\mtaue\propto\mtaud$,
and exhibits reptation--like scaling, $\mtaue\sim N^3$. For tracers which are
much larger than the matrix polymers, which are entangled however,
i.e. $n=N/N_e\gg p\gg 1$, PMC theory finds an intermediate scaling behavior,
$\mtaue\sim N_e^2 p^{3/4} n^{19/8}$. Note that these results, more strongly
than our results for the internal or the center--of--mass dynamics, 
depend on the  approximations made when diagonalizing the equations of motion,
\eqn{1}, where chain end effects have been neglected \cite{fractal}.

 Also, for a PI  tracer in a PBD matrix system--specific 
polymer interactions could strongly influence these results, especially the
parameters characterizing the tracer dynamics. Whereas the parameters, $P_e$,
$\alpha^m$, and $\delta$, specifying the dynamics of the entangled melt
correspond to the ones found (for example)
 from independent viscosity measurements, the
tracer parameters, $N_e$ and $\alpha^{\rm tr}$, must 
differ from the corresponding
values found for the one-component system of tracer polymers. As \eqns{35}{54}
show, these parameters are determined by the tracer--matrix
interactions. Moreover, the importance of the constraint release mechanism
depends on the ratio of the monomeric time scales of the tracer and matrix
polymers.  The corresponding variable is $r=\tau_o^{\rm tr} P_e / \tau_0^m N_e$
in \eqn{64}, where $\tau_0^{\rm tr}$ is the monomeric time scale of the tracer
in the specific matrix it is embedded in, not in its own melt state. 

Adachi and coworkers have observed that the dielectric relaxation times, \taue,
of PI tracers in PBD matrices 
measured in the molecular weight ranges, $-0.2\le \log_{10}n\le 1.45$ and
$0.0\le \log_{10}p\le
2.03$, can be fitted with the   following formula:
\beq{adachi}
\log_{10}(\mtaue/N_e^2\tau_0) = -1.92 + ( 3 - e^{-(P/8500)^{1.5}} ) \;
\log_{10}(M/1900)
\; , 
\eeq
where $\tau_0 N_e^2 = 10^{-3.71} sec$, and the molecular weights of
entanglement are $P_e=1850$ and $M_e=5000$. 
Note that the range  $0.03< \log_{10}p< 1.03$ is of special interest, as there the
effective exponent is found to lie between the expected  asymptotic
behaviors, $\mtaue\sim N^2$ or $\mtaue\sim N^3$. These limits,  Rouse for
$P<P_e$, and  reptation--like  for $P\gg P_e$, are also predicted
by PMC theory.
 For intermediate matrix molecular weights, the PMC approach does not predict simple
power laws as reported \cite{adachi} in \gl{adachi}. A crossover from
$\mtaue\sim N^3$ for $N\ll P$, to $\mtaue\sim N^{19/8}$  for $N\gg P$, for
entangled matrices, $P>P_e$, results in some curvature. The PMC curves matched
to \gl{adachi} therefore do not fit  quantitatively. In the comparison 
of figure \ref{fadachi} the PBD melt is a priori
characterized by parameters, $\alpha=4$
and $\delta=0.05$, found previously in sections 3.A and 3.B. Then three
parameters are varied in the fitting procedure.
For simplicity,
$M_e$ of the PI tracer in the PBD melt is chosen to be the one for a PI melt,
and $\alpha^{\rm tr}=12$ is matched to the asymptotic behavior,
$\mtaue/N_e^2\tau_0 \sim \frac{32}{12} \frac{n^3}{\alpha^{\rm tr}}$. The fitted
 small
value of the time scale ratio, $r=\tau_o^{\rm tr} P_e / \tau_0^m N_e=0.005$,
indicates that the tracer polymer is relatively mobile compared to the matrix
polymers. As is evident from \eqn{64}, the effect of a small parameter $r$ can
also be interpreted as a change in $N_e$, $\Theta \propto r N_e^{-5/2}$, and
without further independent experimental information a closer identification of
the PMC model parameters is impossible. Also,
 the errors introduced into the PMC
results by the neglect of chain end corrections \cite{fractal}
are unknown at present. 
In figure \ref{fadachi},  the reported change \cite{adachi,adachi96}
from Rouse to reptation--like molecular  weight scaling of \taue is
qualitatively correctly reproduced by the PMC results, but
with quantitative deviations. Instead of a scaling exponent which
continuously depends on matrix molecular weight \cite{adachi},
three exponents, $2$, $19/8$ and $3$, are predicted, whose observation depends
on the ratios $N/P$ and $P/P_e$.

Let us recall the predicted behavior of the melt dielectric relaxation time
shown in figure 10 of paper I. There, the dielectric relaxation time of one of
the matrix polymers, i.e. $N=P$ and identical chemistry, was observed to scale
like $\mtaue\sim N^{3.4}$ in an intermediate molecular weight range, 0.5 $\le
\log_{10}(N/N_e)\le $2. Figure \ref{fdilec} shows the dielectric relaxation
times for PI homopolymer melts \cite{adachi1,adachi2,adachi3} and PMC results
for melt like parameters $\alpha=$ 2.2---3.5 and $\delta=0.05$. Both values
agree with the estimates given in section 2 and are compatible with the
analysis of the self diffusion data \cite{fleischer} shown in figure
\ref{fmelt}. Thus, PMC theory also achieves a quantitative description of the
PI melt dielectric relaxation times mainly by adjusting the self--consistent
constraint release mechanism efffects  via the paramter   $\alpha$.  

The molecular weight dependence of the dielectric relaxation time thus is
influenced by the finite size size effects which result from the spatial and
temporal correlations of the matrix constraints. 
The initial decay of the disentanglement process in the end--to--end vector
correlation function, however, is determined by the tracer shape
fluctuations in PMC theory.
 As shown in section 3.A of paper I, the reptation like behavior,
$\varepsilon''(\omega\mtaue\gg1)\sim\omega^{-1/2}$,  emerges from the PMC
description only if 
the internal contributions to the collective tracer dynamics in the RR model
are neglected, which is not justifiable.
 Even in the case of tracer motion in matrices of immobile
polymers, where the reptation--like scaling, $\mtaue\sim N^3$, is predicted,
the tracer shape fluctuations of PMC theory still speed up the initial disentanglement
process, resulting in $\varepsilon''(\omega\mtaue\gg1)\sim\omega^{-x}$, with
$9/32 \le x \le  3/8$ (see section 3.A of paper I). These predictions
qualitatively agree with the  observation \cite{adachi} of ($i$)
frequency power law exponents around
$x=0.22$ to 0.31  for PI tracers  in strongly entangled PBD matrices 
($N\ll P$ and $P\gg P_e$), where $\mtaue\sim N^3$, ($ii$)
exponents $x\approx0.53$ in the Rouse limit, $P_e\ll N\ll P$, and ($iii$) $x$
values in between for matrix molecular weights in between.
Only numerical calculations \cite{ks2,ksneu}
of the full PMC equations can determine whether
quantitative agreement results. In this respect, it is important that the
constraint release correction to the initial decay of the disentanglement
process   can be neglected, as it sets in only for  $t> \mtaud$.

\subsection{Solution Transport Coefficients}

It has been rather well established that the molecular weight dependence of the
viscosity of entangled polymeric solutions is very 
similar, and perhaps identical to within experimental error,
 to the behavior in polymer
melts \cite{pearson87,ferry}. For example, in solutions of PS in a good
solvent, Nemoto and coworkers \cite{nemoto1,nemoto3}  found that for the
semidilute and concentrated solutions the viscosity shows an apparent
power law behavior, $\eta\sim N^{3.5\pm0.1}$ in the range
$0\le\log(N/N_e)\le1.7$. As figure \ref{fvissol} shows, PMC theory can
quantitatively account for these behaviors, and (inverse) entanglement strength
factors  in the physically expected range of
 $\alpha=2.5$ for 40 wt\% PS and $\alpha=4.0$ for 13 wt\% PS are found.
The similarity of the entangled solution viscosity to the melt viscosities
finds a clear physical origin within PMC theory
in the similarity of $\alpha$ for solutions and melts.
From the PRISM estimates, this density and concentration independence of
$\alpha$ has been anticipated. For the semidilute concentration data, 13 wt \%
in figure \ref{fvissol},
it is not clear how to unambiguously compare theory and
experimental data. For low molecular weights, $N\le N_e$, 
the data exhibit a slightly stronger dependence on molecular
weight than explained by the Rouse model, which is assumed in the  PMC approach
to apply
for $N<N_e$. The vertical shift of the 13 wt \% data therefore is unclear.
 We estimate that this is the origin of the variation of
$\alpha$ between 2.5 and 4 found in the fits of figure \ref{fvissol}; these
fits 
result if the experimental viscosities are normalized to $\eta/M=1$ for
$M=M_e$. 

The PMC theory
together with the PRISM estimates of the dynamical parameters explain
that the viscosity versus molecular weight curves for solutions and melts are
similar. Note that the theoretical
predictions for the concentration dependence of $N_e$
can in principle be used, although in our case $N_e$
was taken from experiment \cite{nemoto1,nemoto2,nemoto3}. The similarity 
of solution and melt viscosities results in PMC theory
from the equal influence of
the matrix constraint decay relative to the probe diffusive motion in the
conformational memory function, $M(t)$. Important for this result are two physical
aspects. First, rather long ranged intramolecular correlations dominate
the conformational friction function; see the discussion of $I^M_q$, \eqn{63}. 
This leads to insensitivity of $\eta$ to local length scales like the density
screening length. Second, the (near) cancellation of the concentration
dependence in the PMC entanglement strength factor \cite{ks3},
 $1/\alpha\sim g_d^2 S_0\sim \varrho_m^0$.
 Although at lower solution densities the polymer density
fluctuations, or the osmotic
compressibility, are strongly enhanced (larger $S_0$),
the local contact probabilities of monomers on different polymers,  $g_d$,
diminish, thereby resulting in $\alpha\approx$ constant with varying polymer
concentration.  

When studying the tracer and self diffusion coefficients of the same PS in good
solvent systems \cite{nemoto1,nemoto2,nemoto3}, it is the constraint release
mechanism which leads to the differences between the self, $N=P$, and (infinite
matrix molecular weight) tracer, $P\gg N$, diffusion coefficients. In figure
\ref{fpssol} one observes a similar enhancement of the solution self diffusion
constants relative to the solution tracer diffusivities as has been observed in
figure \ref{fmelt} for the melt diffusion data. There the melt self diffusion
constant at $N/N_e= 4.4$ was a factor 2.1 larger than the asymptotic
expression, \eqn{52}, which neglects finite size and especially constraint
release effects. Due to the similar constraint release parameter,
$\alpha=2.8$, a priori 
chosen in the  analysis of figure \ref{fpssol}, a similar
 enhancement of the self diffusion data (affected by constraint release)
to the $P\gg N$--tracer diffusion constants (unaffected by constraint release)
 is observed. For example, at $n=N/N_e=4.4$ for the 40 wt\% PS
solution, $D(N=P)/D(P\gg N)\approx 2.1$ is found in figure \ref{fpssol}. 

Even though the constraint release mechanism cannot be of relevance to the
(infinite matrix molecular weight) tracer diffusion constants, they do not
exhibit the reptation like scaling, $D\sim N^{-2}$, in  dense PS
solutions \cite{nemoto1,nemoto2,nemoto3}. The experimental data of Nemoto and
coworkers \cite{nemoto1}
which extend to rather high molecular weights, $n\le 70$, exhibit an
apparent power law behavior, $D\sim N^{-2.5\pm0.1}$, and lie well below the
corresponding melt data at high molecular weights plotted in an isofriction
comparison for fixed $n=N/N_e$.
In figure \ref{fpssol} the measured diffusion data are compared to this power
law, to a PMC fit, and to the asymptotic PMC prediction, \eqn{52}
with $\lambda_D=19.2$. The PMC fit uses an independently,  from the viscosity
analysis, determined $\alpha$, and freely varies $\delta$ and $\lambda_D$.
Note that the short dashed curve in figure \ref{fpssol}
shows the asymptotic PMC result including  the Rouse, unentangled behavior,
i.e. $D=D^R/(1+\lambda_D n)$. The enhancement
 of the diffusion, especially of
the tracer diffusion coefficients, relative to the asymptotic behavior, $D\sim
N^{-2}$, therefore cannot result from the crossover to the Rouse law, $D\sim
1/N$. The enhancement of the PS diffusion coefficients in entangled,
PS solutions relative to the short dashed asymptote in figure
\ref{fpssol} is a finite size effect according to the PMC fits
included in figure \ref{fpssol}. 

From the true PMC  asymptote,
$D\sim D^R/ \lambda_D n$, one can deduce that the asymptotic
prefactor, $\lambda_D$, which, in the comparison with the melt diffusion
constants in figure \ref{fmelt}, was found to be around 3 in melts, has to be
appreciably larger in solutions.
 A value of $\lambda_D$ smaller than $\lambda_D=12$ appears
not compatible with the solution data
as they exhibit an increased molecular weight dependence out to the highest $N$
points.  Such a large value of $\lambda_D$ is not
compatible with the relation $\lambda_D=32/3\alpha$,  \eqn{53}, as the value of
$\alpha$ lies in between 2 and 5 as has been discussed in context with the
solution viscosity and the difference of self to tracer solution  diffusion
constants. At present it is not clear how this violation of \eqn{53} can arise
in PMC theory. In the discussion of section 5 we will list possible mechanism.
One possible origin is the simplified treatment of the two--component system,
i.e. polymer solvent mixture, as an effective one--component system. 
Lacking a
quantitative theoretical understanding of $\lambda_D$ in solutions we suggest
using it as a free fit parameter in eqns.  (I.64,I.68,I.69), while
$\alpha=3\pm 1$ may be fixed a priori from independent viscosity measurements. 

The PMC fits to the self and tracer diffusion constants in entangled,
concentrated polymer solutions in figure \ref{fpssol} thus show that again
the apparent power law behavior, $D\sim N^{-2.5}$,
 can be accounted for by a finite size correction 
of PMC theory,  the constraint porosity. It arises from the rather large density
screening length, $\delta=\xi_\rho/b=0.3$, found in the fits. Note that the
constraint porosity corrections, which cause the strong deviations of the
solution diffusivities from the  asymptote, $D\sim N^{-2}$,
in figure \ref{fpssol} have no analog in the reptation/ tube
picture. Consequently, no consistent
 theoretical understanding within that approach has
up to now been presented for the solution data of Nemoto and coworkers
\cite{nemoto} which  explains  the difference \cite{deutschmadden,obukov} of
solution and melt diffusivities and the melt results as well. 

 Within the
PMC theory, the enhanced effective exponent results from the 
gradual build up of the full constraint amplitudes a tracer
polymer feels as it compresses the surrounding entanglement mesh. 
Only when the size of the polymer tracer far exceeds the entanglement length,
$b$, and the compressibility or density screening length scale, $\xi_\rho$,
are the  full matrix constraint amplitudes effective and lead to the
reptation like asymptotic behavior. As the reptation/ tube approach neglects
the spatial variation of the compressibility by simplifying the spatial
correlations of the entanglement constraints to arise solely from the
entanglement length ($\delta=0$ effectively),
 it misses this effect and cannot explain the solution
diffusion coefficients. The PMC approach naturally includes the constraint porosity
effects as it starts from the microscopic correlations of the polymeric liquid
structure. This is especially necessary in polymer solutions where the density
screening length  is not small compared to the entanglement length,
$\delta=\xi_\rho/b=0.3$ is found for the PS solution data of Nemoto et.al.
\cite{nemoto1,nemoto2,nemoto3}.
This length scale ratio is larger in solutions than in melts as has been argued
in section 2. It is rather intriguing, but not understood at present, that the
ratios  of the asymptotic prefactor, $\lambda_D$, in solution to the one in
melts,  and of the length scale ratio $\delta$  in solution to its value in
melts, are rather similar: $\lambda_D^{\rm sol.}/\lambda_D^{\rm melt}\approx 
\delta^{\rm sol}/\delta^{\rm melt}\approx $ 5---6. 

The possibility of a common fit with one set of parameters of self diffusion
relative to its
tracer diffusion analog at three different polymer concentrations in good
solvents shows that the
constraint release enhancement of self to tracer diffusion constants is
correctly captured in PMC theory, and that the PRISM prediction of concentration
independence of the parameters $\alpha$ and $\delta$ in solutions
is accurate. Note that variations of $\alpha$
around $\alpha=3\pm1$ do not worsen the fits. This point is addressed in
 figure \ref{fpssolb} by independent least square fits to the individual data sets;
 $\alpha=2.83$ is apriori fixed.
 Variations of $11\le\lambda_D\le 39$ and
$0.2\le \delta\le0.5$ again emphasize that both parameters are larger for solutions
than for melts, but do not exhibit systematic polymer concentration dependences
in the semidilute and concentrated solution regimes.  

The enhancement of the diffusion coefficients compared to the asymptotic,
reptation like results, $D\sim N^{-2}$, is considerable even for large
molecular weights. For $n=N/N_e=4.4$ and the 40 wt\% PS solution, the tracer
diffusion coefficient
lies a factor 6, and the self diffusion constant a factor 12 above
their common asymptote in figure \ref{fpssol};
 at $n=30$ for the 18 wt\% solution the enhancement
factors are still 2.2 and 1.9, respectively. Extending  the PMC fits, 
a merging of the diffusion data with the asymptote can be expected only for
even larger molecular weights, ca. $n\approx 300$.

\section{Polymer Tracer Motion in Gels: Comparison with Experiment}

As the extension of the PMC theory to polymer tracer diffusion through gels has been
presented in paper I, and since
 tracer motion through gels is the original problem
studied by de Gennes \cite{degrep}, two experimental studies of this problem
are included in this paper. The low field 
mobility, $\mu=D N$, from electrophoresis measurements of
polystyrenesulfonate (PSS) in polyacrylamide (PAA) gels \cite{hoagland} and the
tracer diffusion measurements of PS in polyvinylmethylether (PVME) gels
\cite{lodgegel}, are considered.

The PMC theory can also describe the 
motion of polymer tracers through random materials, where especially gels are
of interest (see paper I). The physical
aspects of constraint porosity in the PMC description naturally 
generalize to fractal media like gels. 
 Two different gel models were mentioned: ``hard gels'', \eqn{76},
which are characterized by rigid density structures which pose the constraints
on the tracer dynamics, and ``soft gels'', \eqn{77}, where the flexibility of
the matrix structure results in much smaller elastic constraints. Simple
extensions of the PMC results for polymer melts and solutions just replace the
tracer and  matrix equilibrium structure by a fractal tracer or gel
structure respectively. Due to the time independence of the arrested gel
structure, the constraint release mechanism is absent in gels or
amorphous solids in general.
 In PMC theory strong enhancements of the tracer diffusivities relative to the
asymptotic power laws are predicted from the fractal generalizations of the
constraint porosity effects.

 As the tracer--gel chemical interactions may show
strong variations, the  possibility to make a priori PRISM estimates for the
dynamical parameters is missing at the present state of the art of PRISM
\cite{kcur2,kcur3}.  From physical considerations, as the
entanglement asymptotic prefactors, i.e. the $\lambda$'s in eqns.
(I.67,I.68,I.79),  clearly are proportional to the tracer matrix interactions,
$\alpha^{-1}\propto \lambda \propto \langle |F_{tg}|^2 \rangle /\varrho_m$ where the ${\bf
F}_{tg}$ are the tracer--gel interaction forces,
and as stronger chemical interactions, like preferential adhesion, clustering
or partial wetting, are expected for polymers in gel pores, 
it can safely be concluded that the asymptotic prefactor $\lambda_g$ in
\eqn{79} may be significantly 
larger than estimated from simple homopolymer solution or melt systems. 

As has been shown in section 6 of paper I, the most important variable
determining the tracer diffusivities is the ratio of gel pore size, $\xi_g$, to
tracer size, $R_g$. Large finite size corrections are predicted by
\eqns{79}{80} for tracers which persist even if $R_g\gg \xi_g$ but are not yet
in the true asymptotic limit, which may require $R_g/\xi_g=10^2$. 
As the asymptotic regime is reached when $R_g\gg\xi_g$, changes in the
mass--size scaling of the tracer polymer do affect the shape of the
pre--asymptotic apparent scaling region. Variations in the gel fractal
dimension, however, mainly 
shift the fitted (unknown) material parameters, but do not significantly
influence the shape of the $D(N)$ curves.

Figure \ref{fhoagl} reproduces the low field
mobility data \cite{hoagland}, $\mu=DN$, of PSS in PAA gels 
of different gel concentrations, $c$. 
From the data in figure \ref{fhoagl} it is apparent that much stronger
molecular weight dependences of $\mu$ are observed than predicted by either
Rouse or reptation theory. This has prompted Muthukumar and Baumg\"artner to
develop the idea of entropic trapping \cite{baummu1,baummu2,baummu3,muthubaum}
as an alternative transport mechanism which may be dominant in the $R_g\approx
\xi_g$ regime..
We will analyze this observation using the finite size effect, namely the
constraint porosity, captured by PMC theory for polymer melts and solutions.

As the PAA gel structure is assumed to
be similar to the solution structure of PAA \cite{hoagland}, the fractal
dimension appropriate for polymers in good solvents, $D_F=5/3$ in Flory
approximation, is chosen in \eqn{75}. Also the soft gel model clearly applies
as the crosslinked polymer matrix does not exhibit significantly
stronger constraints than
the entanglement mesh of the corresponding solution, but only time independent
ones.   We use this consideration also to a priori fix the ratio of 
entanglement length to gel pore size to the value, $\delta=\xi/b=0.3$, found
for solutions in the previous section.  Following the suggestion
 in Ref. 74
the mass size scaling of the tracer polymer is also described by the Flory
behavior, i.e. $\nu=3/5$ in \eqn{74}. 

Figure \ref{fhoagl} includes fits with
the PMC results, \eqns{79}{80}, for the model of  tracer and gel equilibrium
structures described above.
 Least square optimization was used in order to determine the
tracer--gel interaction parameter, $\lambda_g$, and the gel pore size, $\xi_g$,
for three different gel concentrations. It appears physically very reasonable
that the tracer--gel interactions are gel density independent, and
$\lambda_g=54\pm5$ is found for all three densities. Gel pore sizes are
reported in figure \ref{fhoagl}  by stating the PSS degree of polymerization,
$N$, 
giving $R^{\rm PSS}_g(N)=\xi_g$. The expected decrease of the gel pore size
with increasing gel density is clearly observed. A least square fit to the
three data points yields $\xi_g\sim c^{-1.1}$, which 
overestimates the expected 
exponent \cite{hoagland} ($\xi_g\sim c^{-3/4}$ for semidilute good solutions)
and lies closer to the theta solvent values ($\xi_g\sim c^{-1}$ for semidilute
theta solutions). 

In contrast to the expectations of the entropic barrier model of Muthukumar and
Baumg\"artner \cite{muthubaum}, the steepest region of the mobility versus
molecular weight curve is found to lie at $R_g/\xi_g\gg1$. The entropic
barrier model leads to strongly tracer molecular weight dependent
diffusivities when the tracer random coil just fits into the largest gel
pores. Although the PMC constraint porosity effects set in when $R_g\approx \xi_g$, it
extends out to very large tracers and leads to high apparent exponents even when
$R_g\gg\xi_g$. Also in contrast to the entropic barrier model,
the PMC approach includes the smooth crossovers to the
 non--interacting Rouse and the
asymptotic generalized 
 reptative limits. It  not only describes the
region of high effective power laws, but also
describes a smooth variation for all
tracer molecular weights. Note that, for the fractal dimensionalities chosen 
for PSS tracers in PAA gels in figure \ref{fhoagl}, the tracer diffusion
constants do not crossover to the reptation behavior, $\mu=N D\sim 1/N$, at
extremely large tracer molecular weights, but instead exhibit the much weaker
dependence, $\mu\sim N^{-2/5}$, as follows from \eqn{79}. This weak asymptotic
molecular weight dependence is compatible with the data as the crossover to
this asymptote is predicted to take place at much higher (unobservable)
$N$ only. In the
intermediate range effective exponents up to $\mu\sim DN\sim N^{-2.4}$ have
been  reported \cite{hoagland}, which lie in the range and below the
upper bound deduced in paper I for these fractal dimensionalities, $\mu\sim
N^{-2(2-\nu(d-D_F))}\propto N^{-12/5}$.
 Deviations of the fits in figure \ref{fhoagl} for small
molecular weights presumably arise from the inadequacy of the Rouse model for
short polymers in gels.

Another study of the tracer diffusion of PS through PVME gels \cite{lodgegel}
was aimed at studying the problem originally considered by de Gennes
when introducing the reptation concept \cite{degrep}. The gels and their
flexible, neutral  polymer solution precursors were well characterized 
and considerably crosslinked, and should therefore be a  model
system for testing reptation ideas.
 It was suggested that the gels fall  in the light strangulation regime,
where the polymer strands between chemical crosslinks are somewhat shorter
than the ones between the time dependent entanglements of the non--crosslinked
polymer solution. Following the estimates presented in Ref. 75
we assume that for the three different crosslinking densities studied
the length scale ratios, $\xi_g/b$ are somewhat larger than $\delta$ found for
solutions. $\xi_g/b=$ 0.3, 0.4 and 0.5 are chosen as reasonable estimates.
The smallest value, found
for PS solutions, would correspond to the border of the strangulation regime,
 and $\xi_g/b=$ 0.5 to a molecular weight in between crosslinks a factor
of about  2.8 smaller. 

It came as a surprising finding, which was claimed to be in disagreement with
reptation theory, that rather strong molecular weight dependences of the PS
tracer diffusivities in PVME gels were measured, $D\sim M^{-2.7}$ at gel
density $c=0.200 g/mL$ and  $D\sim M^{-2.8}$ at  $c=0.235 g/mL$ respectively
\cite{lodgegel}.
It was stated that the data agree well with diffusion of PS in PVME solution
measurements where larger or equal (in the limit of large matrix molecular
weights)  diffusivities (also showing anomalous exponents) were found
\cite{lodgegel}. Clearly this aspect of the comparison of tracer motion in
solutions and gels is explained by the PMC constraint release mechanism, which
is not present in gels but affects the solution data depending on matrix (and
tracer) molecular weight. Important for this experimental comparison is the
achievement of similar matrix structural correlations in the cases of PVME
solutions and gels \cite{lodgegel}. In PMC theory, the constraint porosity
effect is important  in both solution and gel situations and can explain the 
finite size fluctuation corrections. 

Figure \ref{flodge} shows PMC model
 calculations appropriate for the experimentally
studied system. As argued above, in order  to mimic the gel strangulation regime
the specified values of $\xi_g/b$ are chosen. From the results in figure
\ref{flodge} it can be seen that small variations with $\xi_g/b$ for the other
parameters are obtained. From the stated near
identity of the PVME gel and solution
density structures \cite{lodgegel}, a Flory excluded volume fractal
dimension, $D_F=5/3$, is deduced for the gels.  Screening of the excluded
volume interactions of the tracer is assumed for distances larger than the gel
pore size; i.e. $\nu=1/2$ is chosen. This choice allows us to connect the gel
entanglement strength factor, $\lambda_g$, to the prefactor of the
center--of--mass friction function, $\lambda_D$; i.e. $\lambda_g=\lambda_D/N_e$
follows in \eqn{79}.  In order to crudely
account for the known specific charge transfer attractive
interactions between PS and PVME 
we choose a priori a value $\lambda_g N_e =200$,
which is a factor of about 10 times
larger than the value of $\lambda_D$ found in PS
solutions in section 3.F.
 Changes of the strength factor  $\lambda_g$ by a factor
of 2 would not result in appreciable changes of the results in figure
\ref{flodge}. From the results for polymer tracer diffusion in solutions shown
in figure 7 of paper I, one can estimate that for $\nu=1/2$, solution--like
strength factors could lead to apparent $D$ versus $N$ scaling exponents up to 2.6.

 Most sensitive for the comparison of the PMC constraint porosity
corrections to the measured PS in PVME tracer diffusivities is  the
ratio of tracer size to gel pore size spanned in the experiments. Following the
statement in Ref. 75
we assume the characteristic pore length of the
gel at density $c=0.235 g/mL$ to be $\xi_g=20$\AA, which  agrees with the size
of a PS tracer in good solvent for a PS molecular weight of $M=5000$.
 Note that the experiments span the tracer molecular weight
range $10^4 \le M \le 10^6$, and are therefore
 in the region where $R_g^{\rm tracer} \gg
\xi_g$. For the lower gel density, $c=0.200 g/mL$, we estimate the
PS molecular weight which leads to $R_g^{\rm PS}(M)=\xi_g^{\rm PVME}$ by
assuming excluded volume correlations of the gel, $\xi_g\sim c^{-3/4}$, and of
the PS tracer inside the gel pores, $R_g\sim M^{3/5}$ for $R_g\le \xi_g$; this
leads to $\xi_g(c=0.200 g/mL)$ corresponding to $M^{\rm PS}=5000
(\frac{0.200}{0.235})^{-5/4}\approx 6120$. 

Figure \ref{flodge} shows the PMC tracer diffusion coefficients determined with
these parameters and thus very roughly corresponding to the experimentally
measured PS diffusion coefficients in PVME gels \cite{lodgegel}. Whereas
changes of $\lambda_g$ by a factor 2 and $\xi_g/b$ in the employed range
do not change the results
appreciably, a larger fractal exponent $\nu$ of the tracer chain,
e.g. $\nu=3/5$ if excluded volume for the tracer was not screened, would
decrease the slope of the PMC curves in figure \ref{flodge} down to values
around $D\sim N^{-2.2}$. Changes in the
fractal dimension of the gel only weakly  affect the shape of the PMC curves.
For the chosen parameters, much steeper molecular weight dependences than the
asymptotic, reptation like law, result,
and arise from the spatial correlations of the gel constraints. The tracer
polymer feels the full entanglement constraints only if its size considerably
exceeds the gel pore size and the length scale, $b$, of the elastic
entanglement mesh. For comparison, a power law $D\sim N^{-2.8}$ as reported in
Ref. 75
 is shown in figure \ref{flodge}. The PMC results exhibit
somewhat shallower slopes. In evaluating the agreement with the measured
tracer diffusivities one has to consider the reasonable, but
ad hoc estimates of the parameters,  the experimental uncertainties, and the
complexity of the ternary polymer--solvent--gel system. 

\section{Summary and Conclusions}

The picture of the dynamics of entangled polymeric systems as provided by the
PMC theory shall be summarized in this final section. A comprehensive
overview of all aspects of the PMC approach has been recently given 
\cite{feature}.
The comparisons with experimental data, discussed in the
present paper, show that an unified qualitative and sometimes quantitative,
description of the transport properties of entangled polymer chains has been
achieved. The comprehensive explanation of a wide variety of experimental
observations is noteworthy.  Alternative explanations with phenomenological
theories,  which exist for some of
the observed phenomena, fail to consistently explain all of the studied
 experimental facts. 
The recent measurements of the dielectric relaxation times of tracer polymers
in polymer melts \cite{adachi,adachi96} appear to rule out theoretical models which do
not consider the collective dynamics of the polymer matrix. Explanations of
the anomalous viscosity versus molecular weight scaling, $\eta\sim
M^{3.5\pm0.2}$, provided with models considering a single polymer in a frozen
environment, therefore, appear accidental. The same, but frequency
resolved,  dielectric measurements \cite{adachi,adachi96} stress that  the
rapid initial decay of the disentanglement process (or its shallow high
frequency loss modulus) persists even in the limit where reptation--like scaling
of the final relaxation time is observed. This demonstrates the existence of a
non--reptative effect, unrelated to constraint release,  which determines the primary
disentanglement step. Except for PMC theory,
 where the tracer shape fluctuations are
found responsible, no other consistent theoretical explanation apparently
exists for these
anomalously shallow high frequency wings of the dielectric and shear moduli
spectra. The measurements of tracer diffusion constants in immobile polymer
solutions, i.e. in matrices of  high molecular weight polymers
\cite{nemoto1,nemoto2,nemoto3,nemoto}, and in gels \cite{lodgegel,hoagland},
indicate that strong deviations from reptation--like asymptotes can exist even
if constraint release corrections are absent. The observation of a
$D\sim M^{-2.5}$ scaling in semidilute and concentrated solutions
\cite{nemoto1,nemoto2,nemoto3}, and of $D\sim M^{-2}$ in melts 
\cite{greenkramerlet,greenkramer}, appears to require a proper theoretical
treatment of the polymer liquid structure.

The qualitative picture of the polymer dynamics envisaged by PMC theory
is as follows.
Entanglements arise from the competition of polymer backbone connectivity and
uncrossability due to intermolecular excluded volume forces. Reptation--like
asymptotic transport coefficients are obtained for polymers in fixed dense
environments. This results from the decay of the entanglement friction via the
dynamics of the collective single chain structure factor. Two decay channels,
a coherent (center--of--mass)
 one for small wavevectors, and an internal one, termed tracer shape
fluctuations, for intermediate wavevectors, are associated with
the tracer dynamics
in the friction functions. The tracer shape fluctuations mainly (greatly) speed
up the 
initial disentanglement process compared to pure reptation. 
As entanglement constraints are calculated from the microscopic intermolecular
forces, one material dependent strength factor, $1/\alpha$, enters into the
asymptotic PMC predictions. In contrast to the reptation/ tube approach where
$\alpha$ would be an universal number, in PMC theory
 $1/\alpha$ measures the average,
mean squared intermolecular force per unit density
and therefore is non--universal in
principle. Consequently, non--universal curves for the relative increase of the
transport coefficients, e.g. $\eta/\eta^R$, versus rescaled molecular weight,
$n=N/N_e$, are predicted. From comparison with experiments a rather small
spread 
in the values of $\alpha$ for some polymer melts and solutions is found,
$\alpha=4\pm2$, and the average value lies in the range of the ab initio PRISM
estimate.

Finite size corrections to the transport coefficients result in PMC theory
from both the
decay of the matrix constraints, termed self--consistent
constraint release, and from the
spatial  variations of their amplitudes, called constraint porosity. The
spatial dependence of the amplitudes arises from 
 the elastic entanglement mesh, characterized by an
entanglement length, $b$, and the local compressibility correlations, specified
by the density screening length or physical mesh, 
$\xi_\rho$. Note that in PMC theory these effects
result from straightforward assumptions about the static structure of the
polymeric liquid and its time dependence (see section 3.B of paper I).
 It can be
considered a major achievement of the PMC approach
that the variety of non--Rouse and
non--reptation--like behaviors of entangled polymer dynamics is derived from
one set of microscopically motivated approximations. This approach is very
different from extensions of the reptation/ tube model. There new types of
polymer chain motions are invoked in an empirical and investigator specific
fashion, with parameters sometimes of unclear or vague
 relation to experimental observables,  in order to account for unexpected
(apparent) exponents 
 characterizing  the dependence of a transport coefficient on
macromolecular size. This advantage of PMC theory is accompanied
by the corresponding disadvantage that no intuitive
predictions about the actual monomer trajectories naturally emerges.

An important finding of PMC theory relates to the spatial correlations of the
entanglement friction in the conformational and the center--of--mass motion
\cite{kss2}. In conceptual agreement with the basic reptation/ tube idea, the
center--of--mass motion is determined from rather local friction
contributions. The internal dynamics feels constraints which are spatially
correlated across the tracer size. A first immediate consequence is the
stronger slowing down of the internal dynamics as compared to the translational
diffusion \cite{kss2}. Another consequence are the very different effects the
distinct finite size corrections have on the dynamics. The self--consistent
constraint release
mechanism, which is a more effective entanglement force relaxation mechanism than the tracer
motion for dynamics on the $R_g$ and larger scale,
 is important for friction contributions on
global length scales. The constraint porosity influences the more local friction
contributions most strongly. Therefore, the first finite size effect,
 constraint release, mainly affects the conformational dynamics, i.e. the
internal disentanglement time and the viscosity. The constraint porosity, on
the other hand, accelerates the center--of--mass diffusion and the end--to--end
vector fluctuations of very long tracers. 

The constraint release mechanism is parameterized by the inverse strength
factor, $\alpha$, and therefore is rather independent of concentration and
solvent quality, as follows from the PRISM estimates. Consequently,
PMC theory explains the 
universal observation of apparent viscosity versus molecular weight power laws,
$\eta\sim M^{3.4}$, for both entangled 
polymer solutions and melts. The observation of a
reptation--like scaling of the dielectric relaxation time in matrices of high
molecular weight polymers supports the identification of the PMC constraint
release mechanism as the origin of the stronger increase  of the shear
viscosity. Thus, the failure to observe the asymptotic $N^3$ law is predicted
to be due to the lack of a clear time scale separation between single chain
conformational relaxation and the decay of the entanglement forces. Such a
non--Markovian situation has been invoked in the phenomenological ``coupling
model'' picture of Ngai and coworkers
\cite{ngai1,ngai2}, and is implicit to the ``dynamic cluster''
ideas of Douglas and Hubbard \cite{douglas}.

Spatial correlations of the entanglement constraints do not play an important
role in polymer melts. As stated and shown in figures 5 and 6 of paper I, this
holds in regard to the internal dynamics for any polymer concentration. If
translational diffusion coefficients for fixed reduced degrees of
polymerization are considered, the small density screening length in melts
again leads to negligible constraint porosity effects. In solutions, however,
the center--of--mass mobility is enhanced as the amplitude of the
entanglement constraints are
proportional to the wavevector dependent osmotic 
compressibility. At fixed $n=N/N_e$,
the density screening length is larger relative to the entanglement length 
in solutions than in melts. Smaller effective friction coefficients originate
from the reduced friction contributions on local length scales. 

Crude, a priori  PRISM
estimates for $\delta=\xi_\rho/b$ lie within a factor of $\approx$ 5  from the
experimentally found values.
 In solutions, the magnitude of the uniform PMC
friction  coefficient extracted from fits,
  i.e. $\lambda_D$ in \eqns{52}{68}, deviates from the
prediction, $\lambda_D=32/3\alpha$ \eqn{53}. This deviation is not understood
at present, but possibly results from our simplification of a 
two--component polymer--solvent 
system to an effective, one--component polymeric system. Another
possible origin is the fundamental PMC approximation, \eqn{24}, which
identifies the decay channels of the entanglement friction forces. The
approximation to determine the decay of the entanglement constraints from the
collective tracer structure factor, could be qualitatively correct in solutions
but not quantitatively accurate.
 Note that this approximation is supported by our
findings for polymer melts. There, a consistent set of parameters describes the
translational diffusion coefficient and the viscosity, as shown in section 3.C
of this paper. 

Another physical situation which indicates the importance of the structural
correlations of the matrix entanglements is the polymer  tracer motion through
a gel. It appears as support of the PMC description of the constraint porosity
effects that this common mechanism explains the close connection of tracer
diffusivities measured in entangled solutions and gels \cite{lodgegel}. In our
approach, (soft) gels mainly differ from entangled polymeric matrices in
solutions because of the absence of constraint release contributions. Stronger
constraint porosity effects than in solutions may arise depending on the
fractal dimensionalities of the tracer and the gel, and depending on specific
tracer--gel interactions. Whereas the first effects can again be rationalized
with simple considerations of the ratio of tracer size to gel pore size,
quantitative predictions of the second effects require further experimental and
theoretical work
on polymer--gel interactions. Strong molecular weight dependences of the tracer
diffusivities, e.g. $D\sim M^{-2.5}$ in solutions
\cite{nemoto1,nemoto2,nemoto3}, and  $D\sim M^{-2.8}$ in gels \cite{lodgegel},
result from the enhancement of the center--of--mass motion of tracer polymers
which do not sample the surrounding entanglement constraints uniformly as their
sizes are not yet asymptotically large.

Finally, we conclude with suggestions for new experiments and/ or computer
simulations to further test out theoretical ideas about entanglements in linear
chain melts, solutions, and gels. The key idea is to manipulate (in a
controlled manner) the three fundamental length scales: polymer density
(concentration) screening length, $\xi_\rho$, entanglement mesh length, $b$,
and tracer radius--of--gyration, $R_g$. These nonuniversal structural
parameters have a clear physical meaning, are directly  measurable, and depend
on chemical structure, polymer concentration, temperature, solvent quality, and
external pressure. Significant progress in the ability to test our theory
against existing and future data, and to make ab initio theoretical
predictions, would become possible if there were more direct measurements of
the above quantities. This is especially true in theta solvents and
concentrated solutions (up to the melt) of not only polystyrene, but also many
other materials. For the problem of tracer motion in gels swollen by solvent,
direct measurements of the gel collective structure factor, tracer radius of
gyration, and some reliable knowledge of the effective polymer--gel interaction
are desirable. 

Specific fruitful areas for more experiments include the following. ($i$) Self
and tracer diffusion (especially in the effectively frozen matrix limit) of
polystyrene in good polystyrene solutions in the 40---100 \% polymer
concentration regime. Mapping out the concentration and tracer degree of
polymerization dependences should allow the observation of the crossover from
reptation--like scaling of $D$ in the melt to the ``anomalous'' solution
scaling. Our theory predicts such a crossover, and can be quantitatively 
applied (treating the prefactor $\lambda_D$, which is not understood in
solutions, as a free parameter) if the required structural parameters are
accurately known. ($ii$) Same as point ($i$), but in theta solvents covering
the full semidilute--concentrated--melt range. Since the length scale ratio
parameter $\delta=\xi_\rho/b$ is larger in theta solvents (and more density
dependent), we predict enhanced porosity corrections and larger effective
$N$--scaling exponents for the diffusion constant. ($iii$) Points ($i$) and
($ii$) for different chemical systems. This is important because even at fixed
polymer concentration and solvent quality, the parameter $\delta=\xi_\rho/b$ 
is nonuniversal \cite{raspaud} and can potentially be significantly varied. 
($iv$) More tracer measurements in model gels where the probe and matrix (gel)
polymers are identical. Alternatively, controlled variations of tracer,
solvent, and gel chemical structure could allow the tendency for tracer
adsorption or dewetting from the gel structure to be tuned, which in turn is
predicted to  affect diffusion significantly and in a rationally predictable
manner via the parameter $\lambda_g$. ($v$) Combined self diffusion and shear
viscosity measurements on variable chemical systems in both the melt and
solutions. As usual, the availability of such complementary data allows the
monomer friction constant question to be avoided by studying the product
$D\eta$. More importantly, it places additional constraints on our theory by
requiring the same set of parameters simultaneously explain the
center--of--mass diffusion and conformational (stress) relaxation aspects of
entangled dynamics. ($vi$) Finally, we have discussed elsewhere \cite{fractal}
our suggestions for new experiments at nonzero frequencies, and for transport
coefficients, of entangled fluids composed of macromolecules of variable
fractal dimensionality (e.g. rings) and in variable spatial dimension. Computer
simulation could potentially play a critical role here, although proper
treatment of heavily entangled systems remains in the future.

\bigskip
\bigskip

\leftline{\large{\bf Acknowledgments}}
\bigskip
\noindent 
Partial financial support by the Deutsche Forschungsgemeinschaft under grant Fu
309/1-1, and the United States National Science Foundation MRSEC program via
grant number  NSF-DMR-89-20538,  are gratefully acknowledged.

\newpage
\leftline{{\large{\bf Table:}} Material Parameters of the Different Polymer
Systems Studied}
\begin{table}
\begin{tabular}{|*{10}{c|}}
\hline 
Polym. &
PBD &
PBDh &
PDMS &
PEO &
PI &
PS &
PS &
PS &
PS \\ 
\hline
$\frac{\mbox{\small melt}}{\mbox{\small sol.}}$ & 
m  & m & m & m & m & m & 13 wt \% & 18 wt \% & 40 wt \%\\
\hline
$M_e$ &
1.85 10$^3$ & 1.24 10$^3$ & 1.2 10$^4$ & 2.2 10$^3$ &5.0 10$^3$ & 1.8 10$^4$ & 
1.3 10$^5$ & 9.6 10$^4$ & 4.23 10$^4$\\
\hline
$N_e$ &
34 & 44$^e$  & 162 & 50 & 74 & 173 & 
1250 &  923 & 407 \\
\hline
$\alpha$ & 4.9$^a$, 3.3$^b$ & 4.3$^c$ & 2.8$^b$ & 2.6$^b$ & 2.2$^b$, 3.5$^d$ & 2.2$^b$
& 2.8$^b$ & 2.8$^b$ & 2.8$^b$\\
\hline
$\delta$ & 0.01 & 0.03 & 0.0 & 0.05 & 0.05 & 0.05 & 0.4 & 0.3 & 0.4\\
\hline
$b$ (\AA) & 44.4 & 33.9 & 78.6 & 37.5 & 62.0 & 76.5 & & & \\
\hline
$\xi_\rho$ (\AA) & 2.30 & 1.75 & 4.06 & 1.94 & 3.20 & 3.95  & & & \\ 
\hline
\end{tabular}
{\small $^a$ from viscosity; $^b$ from diffusion; $^c$ from viscosity and
diffusion; $^d$ from $\tau^\epsilon$; $^e$ 
$M_0^{\rm PBDh}=28\, $ used.}
\bigskip
\end{table}

\bigskip

\noindent
Molecular weights of entanglement, $M_e$, are taken from the experimental
studies whose data are analyzed, and entanglement degrees of polymerization,
$N_e$,  are calculated from them using monomer repeat masses \cite{flory}; 
entanglement lengths, $b$,  are taken from Ref. 46
 for $T=413$ K.  The packing lengths \protect\cite{fetters}, $p$,
are identified with the density screening
length, $\xi_\rho\approx1/\varrho_m\sigma^2$, and are also included.
Melt or solution case are indicated by m or by the weight concentration of the
polymeric component. Inverse strength parameters, $\alpha$, and length scale
ratios, $\delta$, are determined by the fits shown in figures 1 --- 4, 7 and 8.

\begin{figure}[H]
\centerline{\rotate[r]{\epsfysize=16.cm
\epsffile{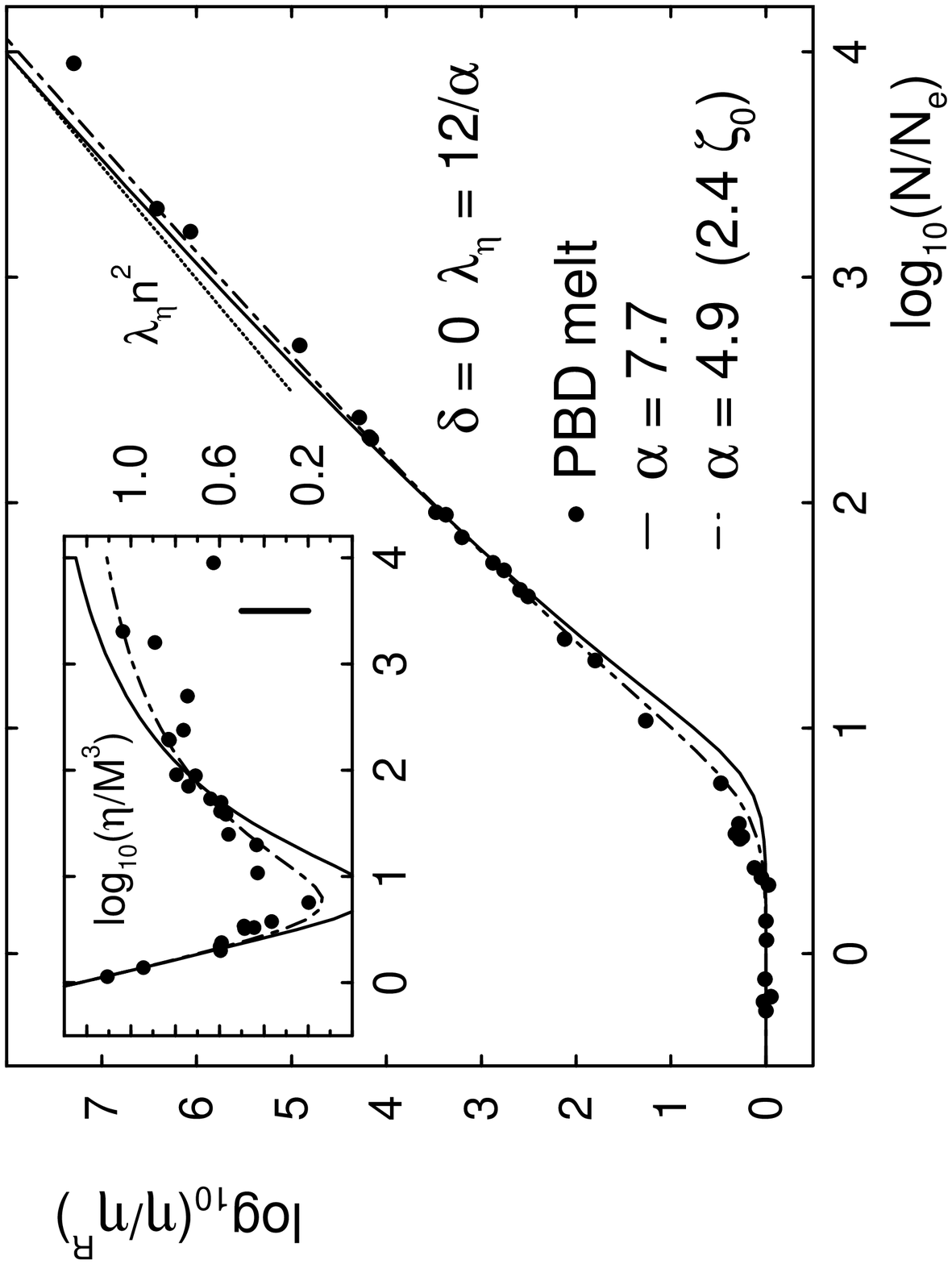}}}
\caption{ }\label{fpbd}
\end{figure}

\begin{figure}[H]
\centerline{\epsfysize=16.cm 
\epsffile{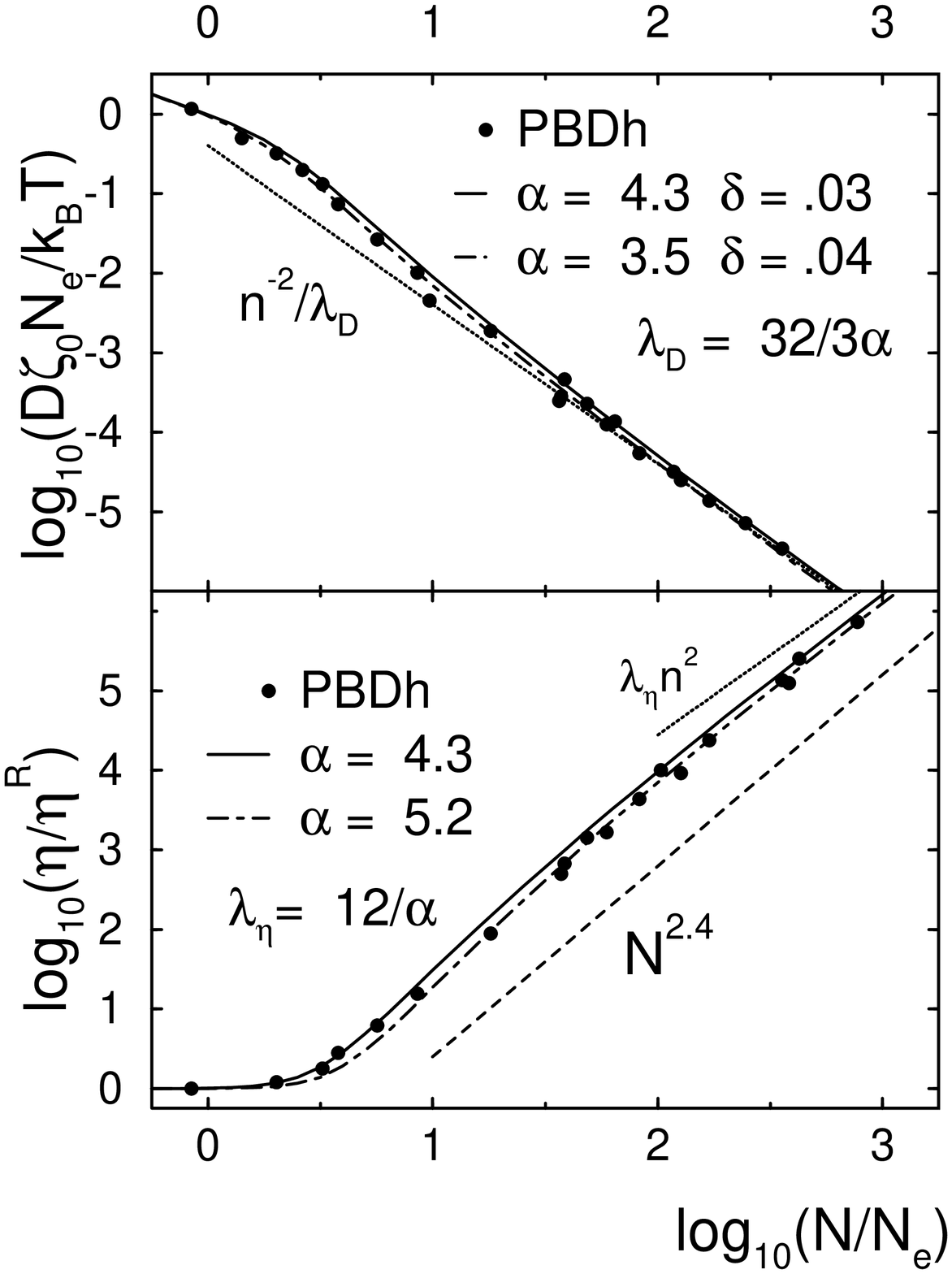}}
\caption{ }\label{fpbdh}
\end{figure}

\begin{figure}[H]
\centerline{\epsfysize=16.cm 
\epsffile{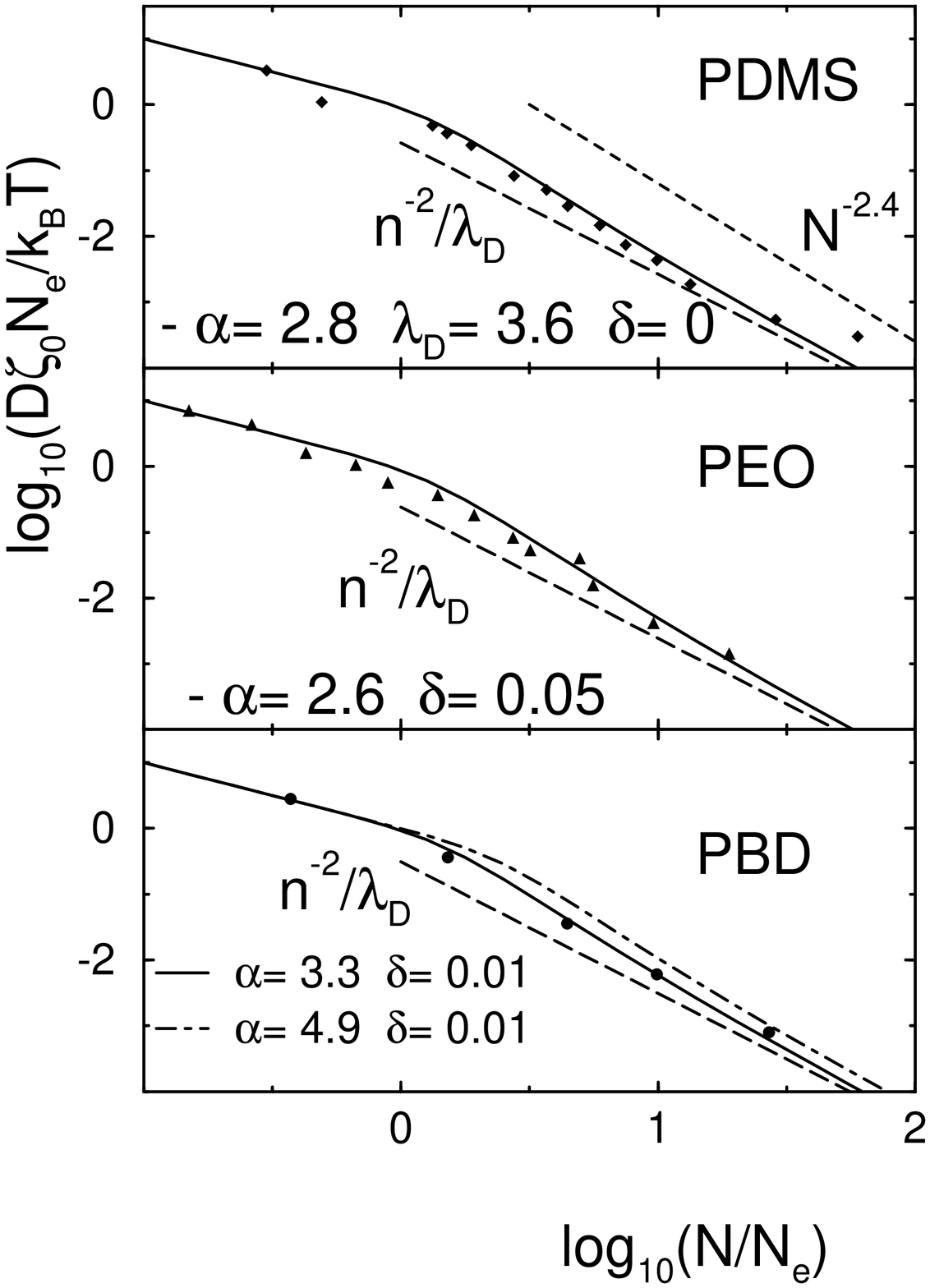}}
\caption{ }\label{fmeltb}
\end{figure}

\begin{figure}[H]
\centerline{\rotate[r]{\epsfysize=16.cm 
\epsffile{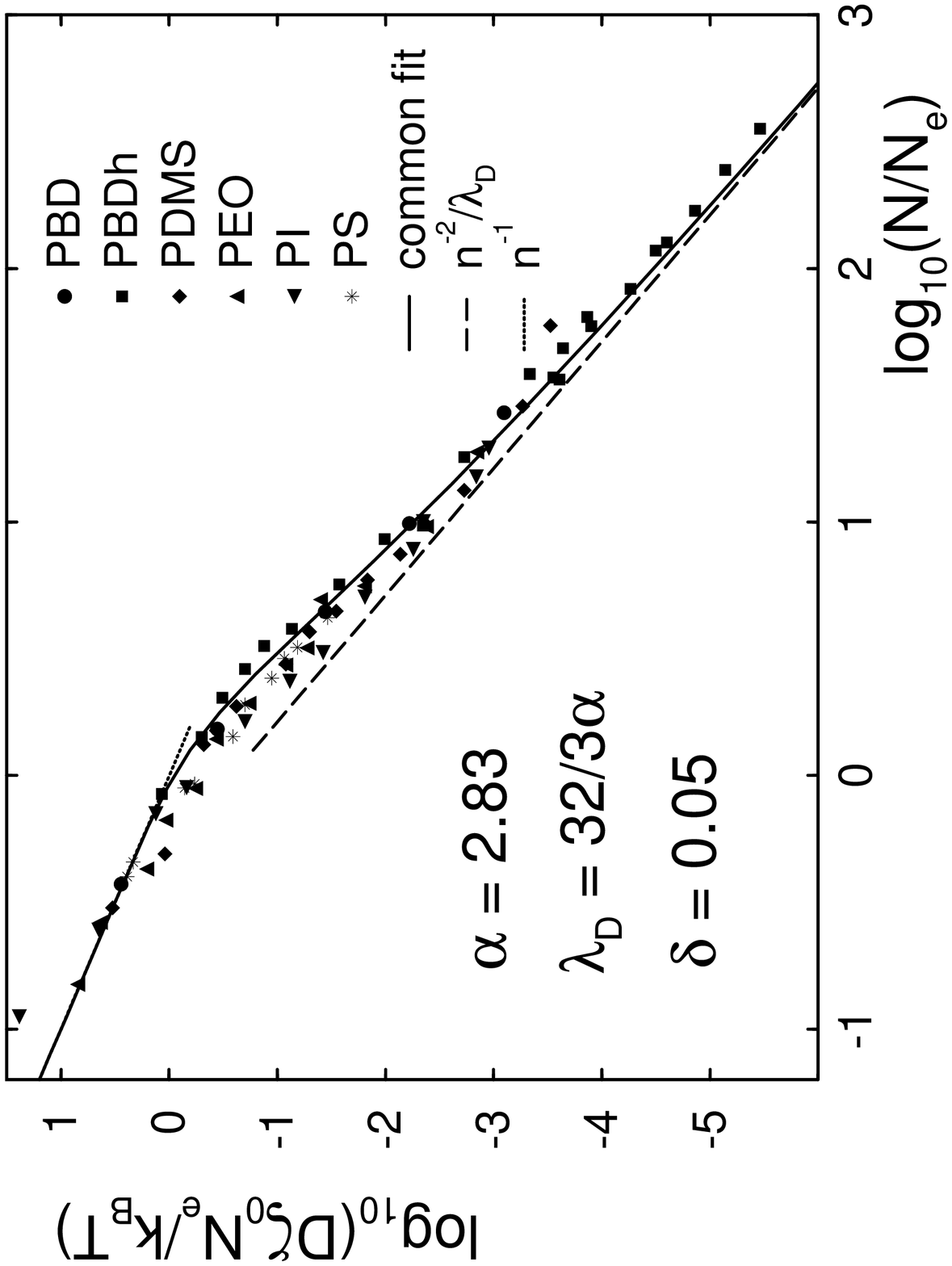}}}
\caption{ }\label{fmelt}
\end{figure}

\begin{figure}[H]
\centerline{\rotate[r]{\epsfysize=16.cm 
\epsffile{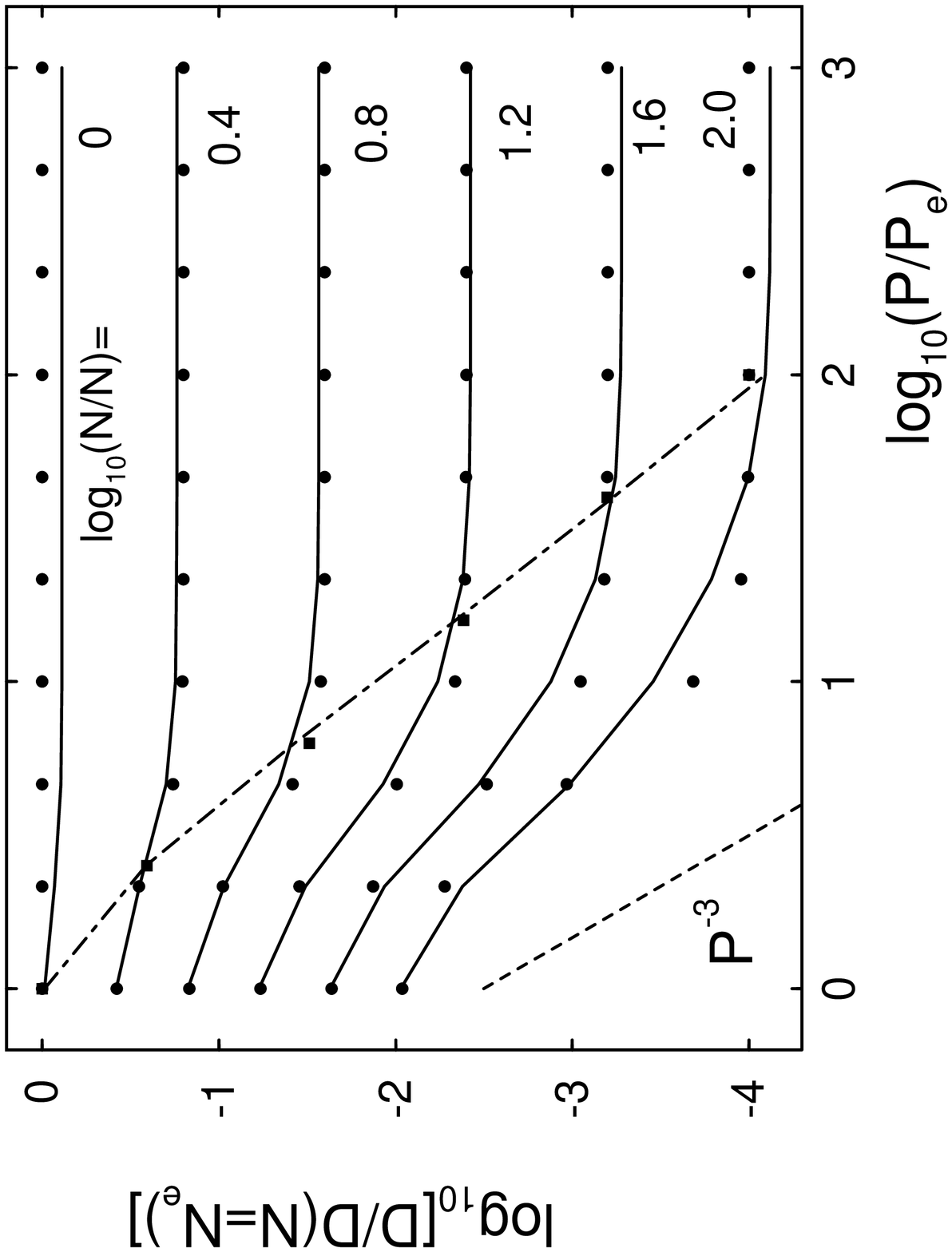}}}
\caption{ }\label{fgrass}
\end{figure}

\begin{figure}[H]
\centerline{\rotate[r]{\epsfysize=16.cm 
\epsffile{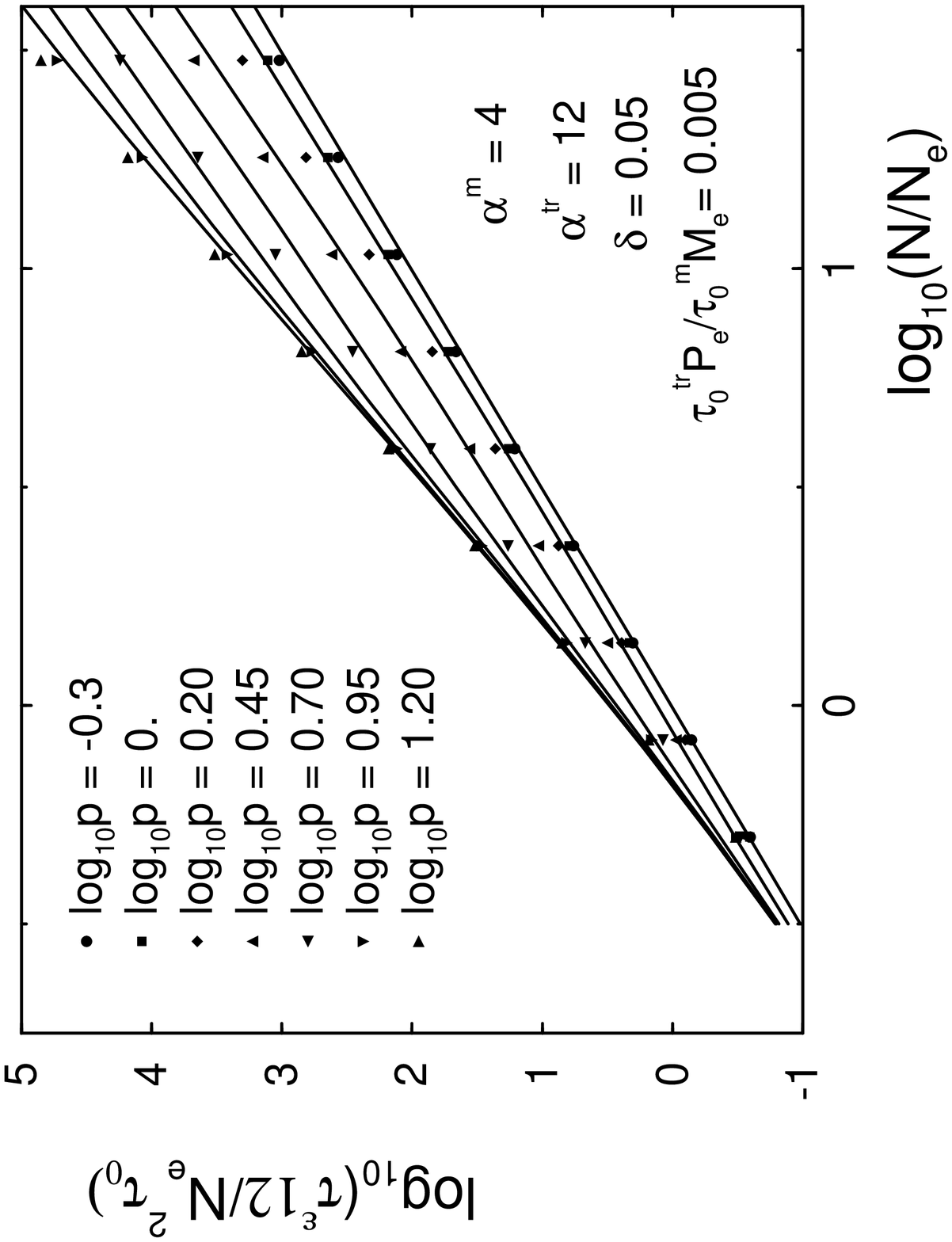}}}
\caption{ }\label{fadachi}
\end{figure}

\begin{figure}[H]
\centerline{\rotate[r]{\epsfysize=16.cm 
\epsffile{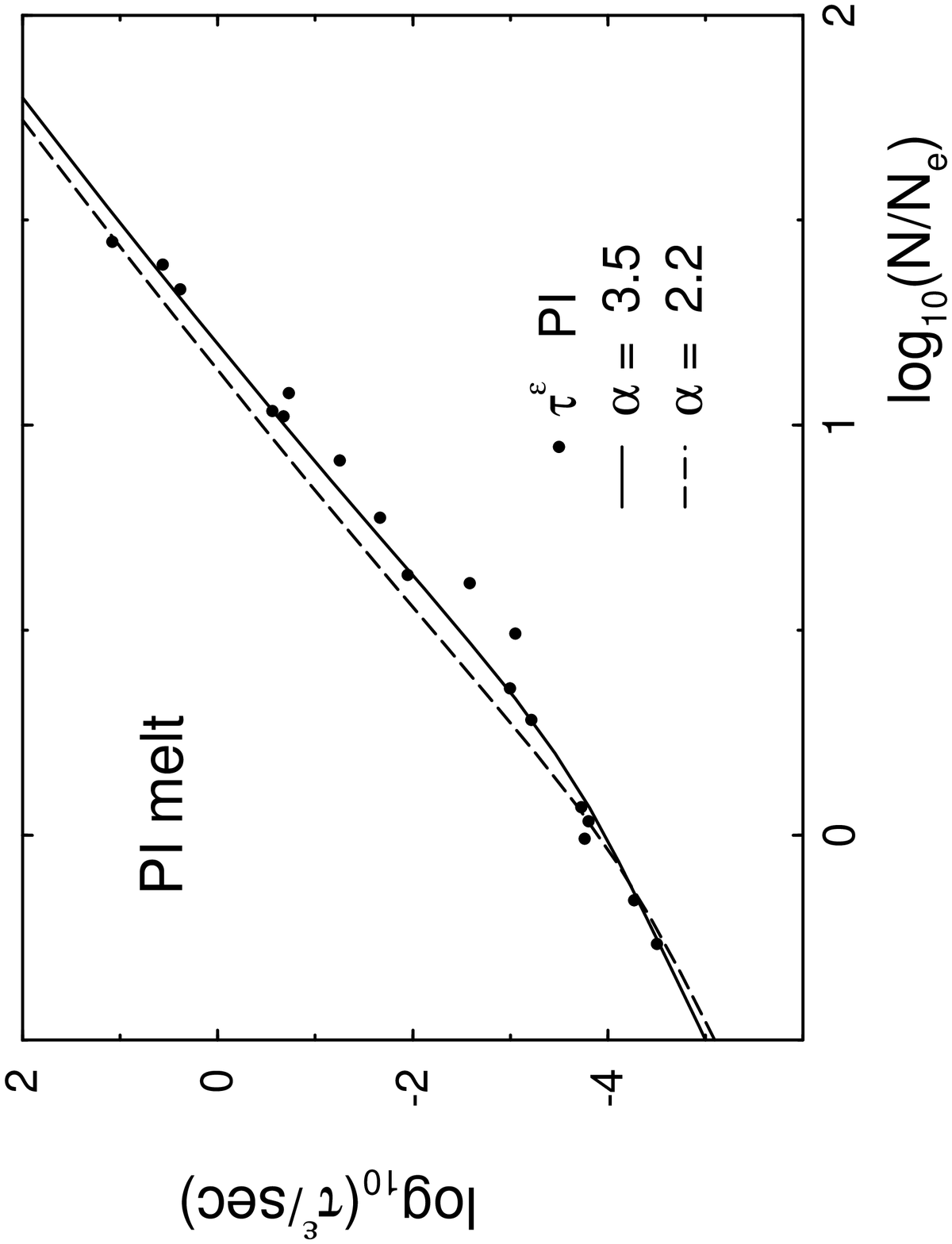}}}
\caption{ }\label{fdilec}
\end{figure}

\begin{figure}[H]
\centerline{\rotate[r]{\epsfysize=16.cm 
\epsffile{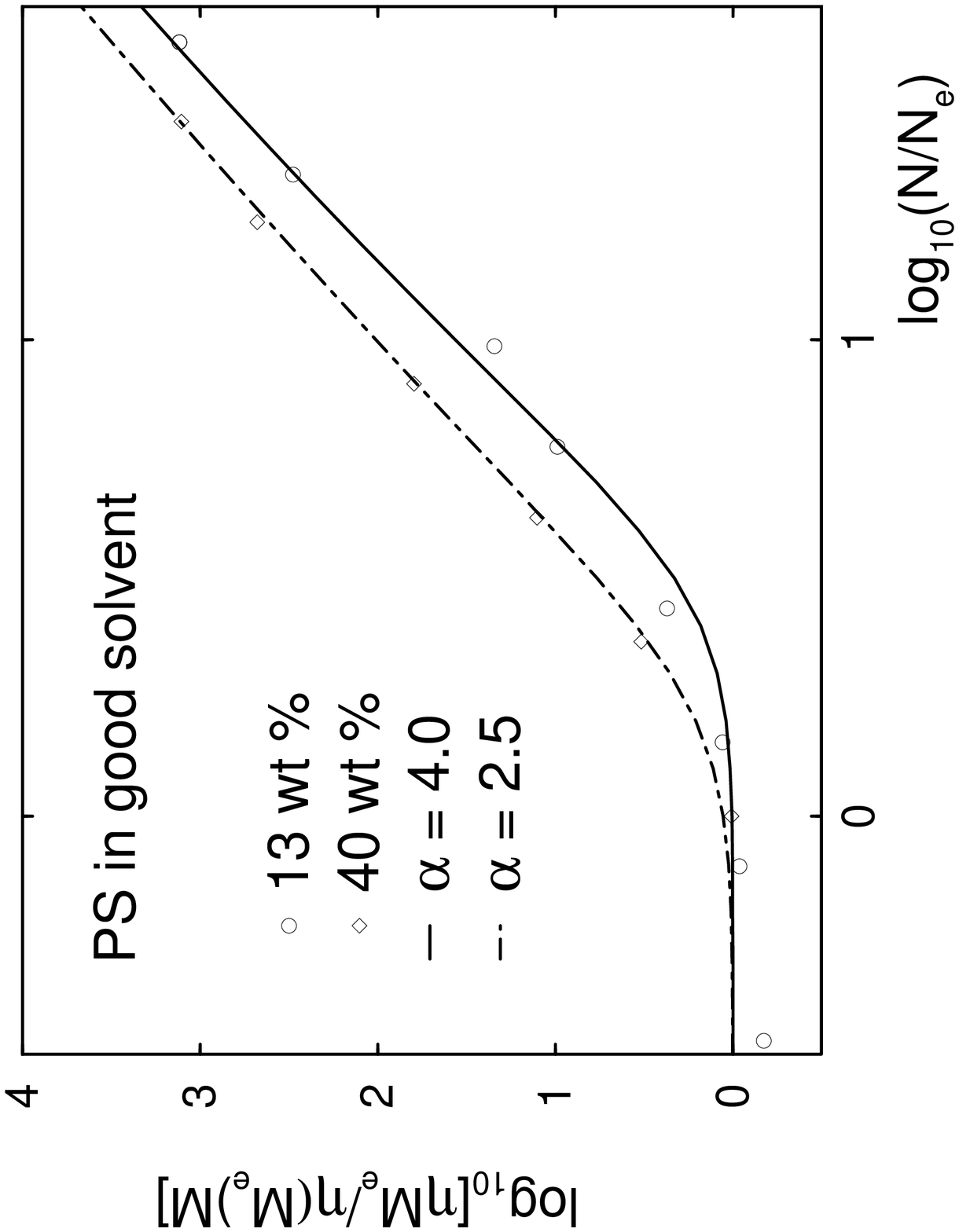}}}
\caption{ }\label{fvissol}
\end{figure}

\begin{figure}[H]
\centerline{\rotate[r]{\epsfysize=16.cm 
\epsffile{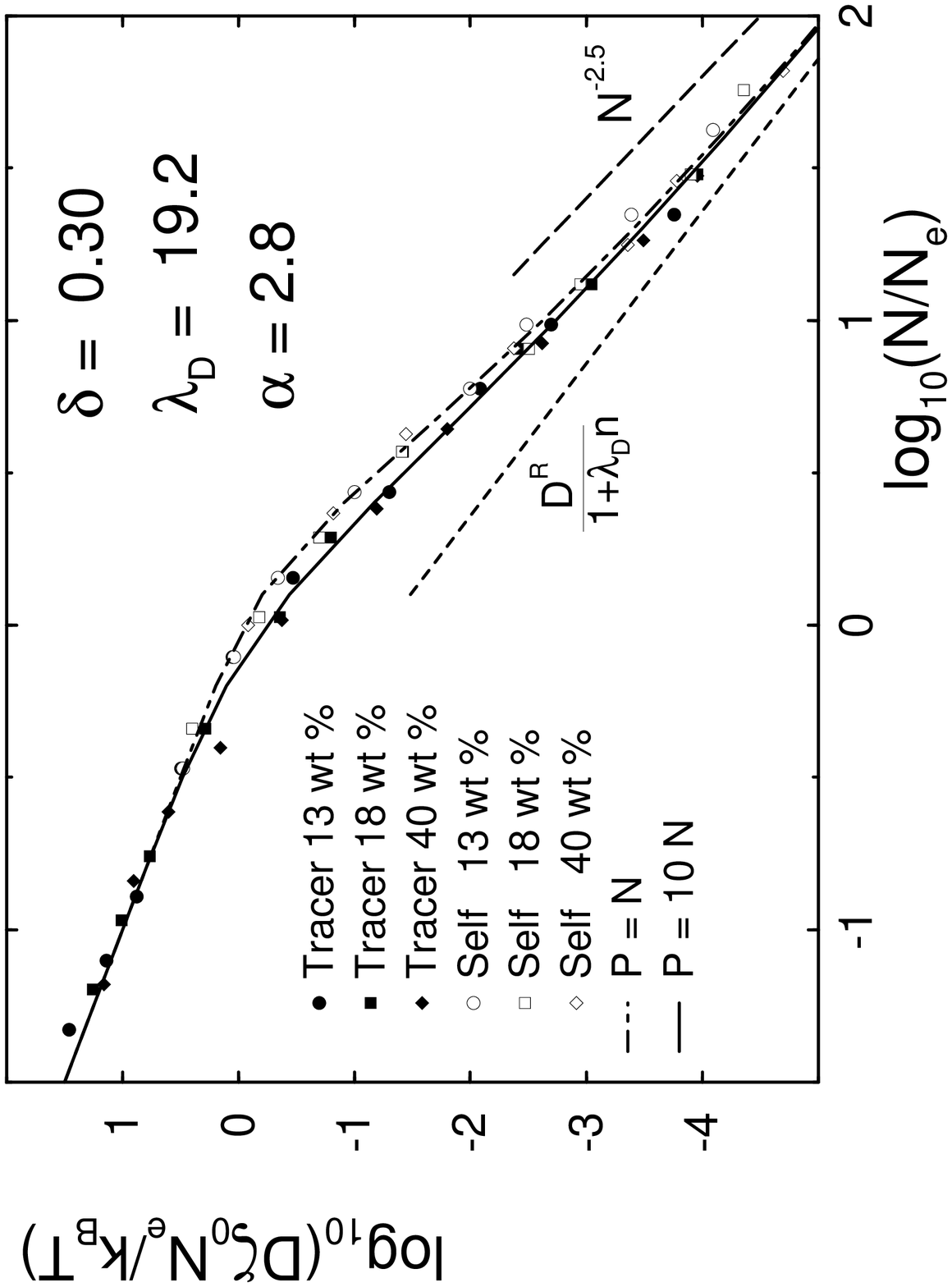}}}
\caption{ }\label{fpssol}
\end{figure}

\begin{figure}[H]
\centerline{\epsfysize=16.cm 
\epsffile{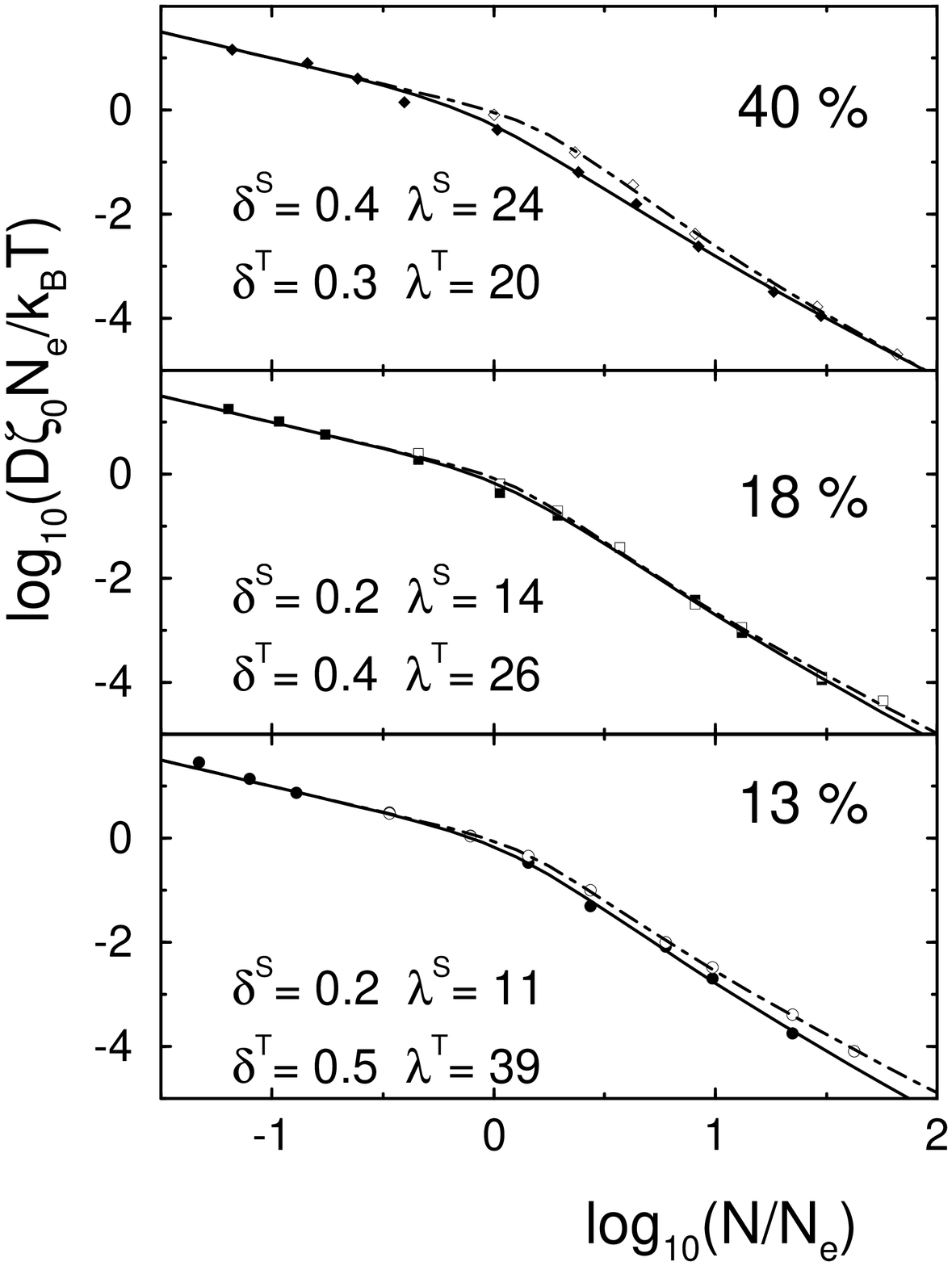}}
\caption{ }\label{fpssolb}
\end{figure}

\begin{figure}[H]
\centerline{\rotate[r]{\epsfysize=16.cm 
\epsffile{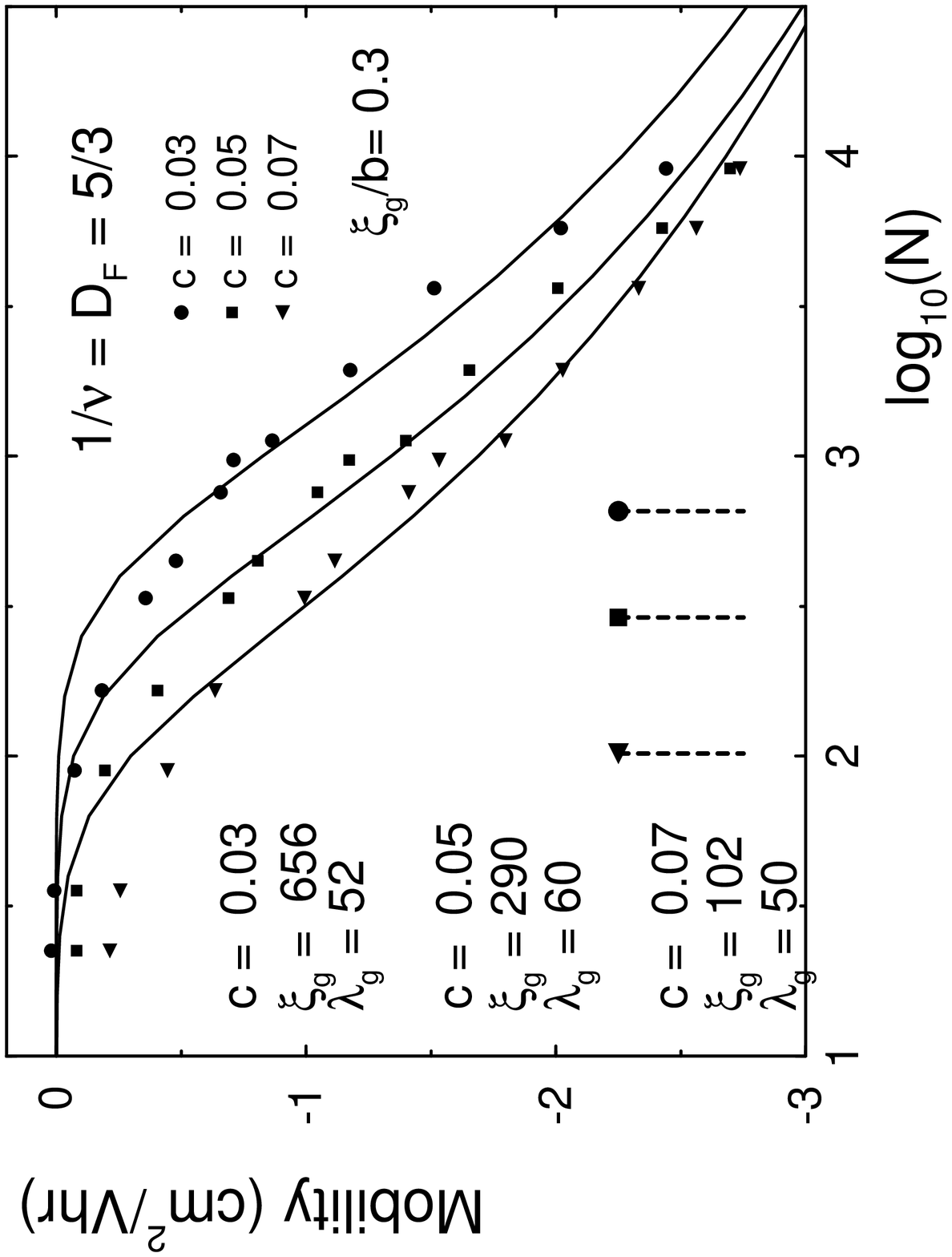}}}
\caption{ }\label{fhoagl}
\end{figure}

\begin{figure}[H]
\centerline{\rotate[r]{\epsfysize=16.cm 
\epsffile{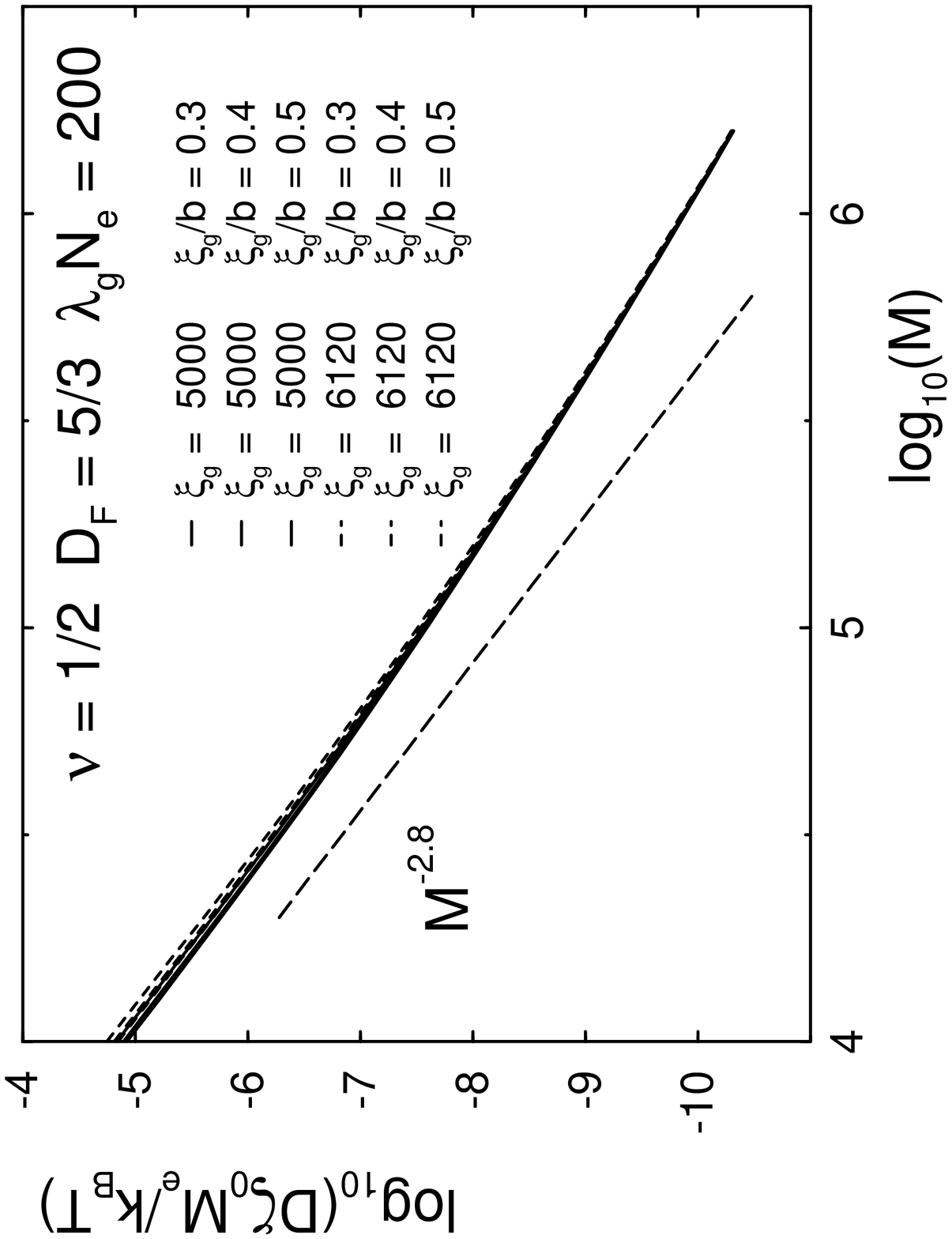}}}
\caption{ }\label{flodge}
\end{figure}

\newpage
\bigskip
\bigskip
\leftline{\large{\bf Figure Captions}}
\begin{itemize}
\item{\bf Figure 1:}$\;$ Reduced viscosity, $\eta/\eta^R$,
for a melt of polybutadiene  \cite{colby} (PBD). The
solid line is a one--parameter
 PMC fit using $\alpha=7.7$, the chain curve corresponds to
$\alpha=4.9$, including the correction for the monomeric friction coefficient,
$\zeta_0^{\rm CFG}=2.364 \zeta_0$, suggested in Ref. 5.
The dotted line
is the asymptote for the second fit. The inset shows $\log_{10}{\eta/M^3}$
shifted vertically by a constant (10) versus logarithm of 
 reduced molecular weight.
The lines correspond to the fits in the main figure. A vertical bar denotes the
error bar reported in Ref. 61
arising from uncertainties in the 
molecular weight and viscosity measurements. 

\item{\bf Figure 2:}$\;$ Self diffusion constant ($D$, upper figure) and
viscosity ($\eta$, lower figure) versus reduced molecular weight for a melt of
hydrogenated polybutadiene \cite{pearson}  (PBDh).
 The solid lines correspond to
fits to $D$ and $\eta$ with a common parameter $\alpha=4.3$; additionally
$\delta=0.03$ is found for the diffusion constant.  The dotted lines are the
asymptotic behaviors. The chain curves show the
best independent fits leading to slightly different $\alpha$ (and $\delta$ for
$D$).  The viscosity data are compared to a  power law $N^{2.4}$ (short
dashes).

\item{\bf Figure 3:}$\;$ Self diffusion constants in melts of 
polydimethylsiloxane (PDMS), polyethyleneoxide \cite{appel} (PEO), and
polybutadiene \cite{fleischer} (PBD) 
versus reduced molecular weight. PMC fits to the individual data sets are shown
using the denoted parameters. For PEO and PBD, the relation
$\lambda_D=32/3\alpha$, \eqn{53}, is enforced. For PDMS,
a unconstrained fit, varying $\alpha$, $\lambda_D$ 
and $\delta$ is shown as solid line, and
results in a small deviation of $\lambda_D$ from its theoretical value,
$32/3\alpha\dot=3.8$. A power law $D\sim M^{-2.4}$ (short dashes) is compared
to the PDMS data.
The high molecular weight asymptotes are denoted by  long dashed
lines. The dot--dashed curve is a fit to the PBD data with the best
$\alpha=4.9$  from the viscosity analysis  of figure \ref{fpbd}. 

\item{\bf Figure 4:}$\;$ Self diffusion constants for polymer melts versus
reduced molecular weight. A common fit for the data of different polymers  with
the indicated parameters  is shown as a solid line. The
asymptotes, Rouse and reptation--like, are shown with prefactors corresponding
to the fit. 

\item{\bf Figure 5:}$\;$ Tracer diffusion coefficients versus matrix molecular
weight, $p=P/P_e$, for different tracer molecular weights, $n=N/N_e$, as
specified in the figure. The model of Grassley \cite{grassley},
eqn. (\protect\ref{grass}), which has been fitted to polystyrene (PS)
 melts by Green and
Kramer \cite{greenkramer} is shown by
circles for $n\ne p$ and by squares for $n=p$.
The PMC fit is shown by the solid curves; parameters are $\alpha=3$ (a priori fixed),
$\delta=0.07$  and $\lambda_D=1.58$. The corresponding self diffusion
coefficients are indicated by a chain curve.

\item{\bf Figure 6:}$\;$ Dielectric relaxation times, \taue according to \gl{adachi},
of a polyisoprene tracer \cite{adachi} versus reduced tracer molecular weight, $N/N_e$, for
the indicated various polybutadiene matrix  molecular weights, $p=P/P_e$.
The lines correspond to PMC  results from eqn. (I.70)  
with the parameters specified in the figure and the Table.

\item{\bf Figure 7:}$\;$
Dielectric relaxation times for  polyisoprene (PI) melts \cite{adachi3}
versus reduced molecular weight. The PMC curves are calculated for
the parameters $\alpha=3.5$ (solid line) and $\alpha=2.2$ (dashed line), where
the latter value is taken from the fit to the PI self diffusion constants shown
in figure 4 and exhibits large uncertainties; the a priori chosen melt--like value
$\delta=0.05$ is used. 

\item{\bf Figure 8:}$\;$ Viscosities of  solutions of 13 wt\% (circles) and 40
wt\% (diamonds) polystyrene in a good solvent \cite{nemoto1,nemoto3}. The PMC
fits use 
the parameters $\alpha=4.0$ (solid line) for the 13 wt\% and $\alpha=2.5$
(chain curve) for the 40 wt\% data.

\item{\bf Figure 9:}$\;$ Self (open symbols) and tracer (filled symbols)
diffusion constants for solutions of polystyrene (PS) in dibutyl phthalate
(good 
solvent). The reported power law, $D\sim N^{-2.5}$, is indicated by a long
dashed line \cite{nemoto1,nemoto2,nemoto3}. A common fit to the self (chain
curve, identical tracer, $N$, and matrix polymer, $P$,  degree of
polymerization) and tracer (solid line, $P=10 N$) diffusion data for
the three densities is shown  and differs in the constraint release
contribution only;  a strong overshoot compared to the asymptote without finite
size effects, $D=D^R/(1+\lambda_D n)$ (short dashes), arises from the
constraint porosity. The theoretical connection of  $\alpha$ and $\lambda_D$ is broken. 
 
\item{\bf Figure 10:}$\;$ Independent PMC fits to the self ($S$; open symbols) and
tracer ($T$; filled symbols) diffusion constants for solutions of polystyrene (PS)
in dibutyl phthalate (good  solvent) from \cite{nemoto1,nemoto2,nemoto3}; 
$\alpha=2.8$ is kept fixed for all fits, and the found parameters $\delta$ and
$\lambda_D$ are listed in the figure.
Error bars for $\delta$, $\delta=0.33\pm0.12$, and $\lambda_D=22\pm10$
follow. 

\item{\bf Figure 11:}$\;$ Low field mobilities of polystyrenesulfonate (PSS)
tracers
in three polyacrylamide (PAA)  gels of different concentration $c$
(wt. fraction) versus tracer degree of polymerization,
 from Ref. 74.
 The full lines are PMC fits with the ``soft gel'' model,
eqns.  (I.77,I.79,I.80),  with the parameters denoted; gels pore
sizes, $\xi_g$, are recorded by stating the degree of polymerization
  of a PSS tracer
with identical radius of gyration. The vertical lines mark the tracer degrees
of polymerization where $R_g(N)=\xi_g(c)$.

\item{\bf Figure 12:}$\;$ Tracer diffusion constants of polystyrene (PS)
as a function of tracer molecular weight ($M$)  in 
polyvinylmethylether (PVME) gels for parameters comparable to the study in
Ref. 75
calculated from the ``soft gel'' model, eqns.  (I.77,I.79,I.80).
Gel pore sizes, $\xi_g$,  are denoted by the molecular weights of PS tracers
with identical radii of gyration. 
 Ratios of pore size to entanglement length, $\xi/b$,
are used appropriate for polymer solutions, $\xi/b\approx 0.30$,  and lightly
strangulated gels, $\xi/b$ larger.

\end{itemize}

\newpage
\pagestyle{empty}

\newpage

\newpage

\newpage

\newpage

\newpage

\newpage

\newpage

\newpage

\newpage

\newpage

\newpage

\newpage

\end{document}